\documentclass[fleqn,usenatbib]{mnras}
\usepackage{newtxtext,newtxmath}
\usepackage[T1]{fontenc}
\usepackage{ae,aecompl}

\usepackage{graphicx}	
\usepackage{amsmath}	
\usepackage{hyperref}
\usepackage{xcolor}

\newcommand{\gr}{\textsuperscript{0.1}($g-r$)}
\newcommand{\Mhr}{$M_r^h$}
\newcommand{\Mhrl}{\textsuperscript{0.1}{\textit{M}}\textsubscript{\textit{r}}~-~5 log~\textit{h}}
\newcommand{\vmax}{v\textsubscript{max}}
\newcommand{\vpeak}{v\textsubscript{peak}}
\newcommand{\vform}{v\textsubscript{form}}
\newcommand{\zform}{$z$\textsubscript{form}}
\newcommand{\zapp}{$z$~$\sim$~}
\newcommand{\grcut}{\protect{^{0.1}(g-r)_{\textup{cut}}}}

\title[The Rosella mock]{Rosella: A mock catalogue from the P-Millennium simulation}

\author[Safonova, Norberg \& Cole]{
Sasha Safonova,$^{1, 2, 3}$\thanks{E-mail: sasha.safonova@yale.edu}
Peder Norberg,$^{2, 4, 5}$
Shaun Cole$^{2}$
\\
$^{1}$Department of Astronomy, Yale University, New Haven, CT 06520, USA\\
$^{2}$Institute for Computational Cosmology, Dept. of Physics, Univ. of Durham, South Road, Durham, DH1 3LE, UK\\
$^{3}$NSF Fellow\\
$^{4}$Centre for Extragalactic Astronomy, Dept. of Physics, Univ. of Durham, South Road, Durham DH1 3LE, UK\\
$^{5}$Institute for Data Science, Dept. of Physics, Univ. of Durham, South Road, Durham DH1 3LE, UK
}

\date{Accepted XXX. Received YYY; in original form ZZZ}

\pubyear{2020}

\begin{document}
\label{firstpage}
\pagerange{\pageref{firstpage}--\pageref{lastpage}}
\maketitle

\begin{abstract}
The scientific exploitation of the Dark Energy Spectroscopic Instrument Bright Galaxy Survey (DESI BGS) data requires the construction of mocks with galaxy population properties closely mimicking those of the actual DESI BGS targets. We create a high fidelity mock galaxy catalogue, including information about galaxies and their host dark matter subhaloes. The mock catalogue uses subhalo abundance matching (SHAM) with scatter to populate the P-Millennium N-body simulation with galaxies at the median BGS redshift of $\sim$ 0.2, using formation redshift information to assign \gr \ rest-frame colours. The mock provides information about $r$-band absolute magnitudes, \gr \ rest-frame colours, 3D positions and velocities of a complete sample of DESI BGS galaxies in a volume of (542 Mpc/$h$)\textsuperscript{3}, as well as the masses of host dark matter haloes. This P-Millennium DESI BGS mock catalogue is ideally suited for the tuning of approximate mocks unable to resolve subhaloes that DESI BGS galaxies reside in, to test for systematics in analysis pipelines and to interpret (non-cosmological focused) DESI BGS analysis. 
\end{abstract}




\section{Introduction}

Upcoming cosmological surveys, such as the Dark Energy Spectroscopic
Instrument (DESI) survey\footnote{\url{https://www.desi.lbl.gov/}} \citep{DESI2016, desiCosmoSim2018},
Euclid\footnote{\url{https://www.euclid-ec.org/}} \citep{Laureijs2011}, LSST\footnote{\url{https://www.lsst.org/}} \citep{Ivezic2008}, the Subaru Prime Focus Spectrograph (PFS)\footnote{\url{https://pfs.ipmu.jp/index.html}} and WFIRST\footnote{\url{https://wfirst.gsfc.nasa.gov/index.html}} aim to map the cosmic structures with the goal of measuring the structures' growth, distribution and the expansion history of the Universe. Cosmological surveys enable measurements of galaxy clustering, redshift-space distortions and weak lensing, among other qualities of the Universe. These measurements can constrain theories behind cosmic acceleration \citep{Efstathiou1990, Riess1998, Perlmutter1999}, test general relativity, and give us greater insight into the nature of dark matter.

The often used way of extracting information from such surveys is to compare summary statistics between observed data and mock data generated from theoretical predictions \citep[e.g.][]{DeRose2019, Smith2020}. In order to compare theoretical predictions to observed quantities, we must create a medium that renders both sides of scientific endeavor --- theory and experiment --- directly comparable. In the context of cosmology and the large-scale structure of the Universe, that medium is a mock catalogue. Such a catalogue serves as a container of data about the quantities we could feasibly observe with cosmological surveys. These quantities might include the masses of galaxies or their brightnesses (in single or multiple bands), galaxy positions, velocities, redshifts, spectra, object type and more.

To be a useful connector of theory to observations, mock data must provide quantities that resemble the observations against which it will be compared. The quantities should satisfy two major requirements. First, the mock quantities must be statistically equivalent to real quantities on the level of individual objects. This can be achieved by, for instance, connecting theoretical predictions with empirical measurements from past surveys. 

The mock data's large-scale structures, as well as its summary statistics, should closely resemble what we observe in the local Universe. Were our simulations and mock data produced from a model that perfectly represented the Universe, the mock data we create from simulations should be indistinguishable from observed data if we examined both side-by-side. This level of statistical resemblance enables cosmologists to make comparisons between theory and observations at high levels of accuracy. 

Mock catalogues can be used to develop and test the analysis tools intended for completed and upcoming surveys because a mock's cosmology is known a priori. The value of a number of parameters of interest can be measured directly in a mock, without the assumptions that are necessary in analyses of real data. Cosmological surveys also require mocks for testing observational strategies and quantifying biases \citep[e.g.][]{Smith2017}.

Modern cosmological surveys, such as eBOSS \citep{Blanton2017, Dawson2013}, DESI \citep{DESI2016}, and LSST \citep{Ivezic2008}, require simulations that cover volumes that exceed 100 [Gpc/$h$]$^3$ in a multitude of realisations. Such great volumes are motivated by a combination of the scientific questions that the surveys attempt to tackle, as well as the systematics that accompany real-world observations.

For instance, for the analysis of systematics for measurements of baryon acoustic oscillations (BAO), volumes of the order of \protect{$\sim 200$ [Gpc/$h$]$^3$ are necessary \citep{desiCosmoSim2018}}. The simulations tailored for such measurements should cover volumes that are at least ten times greater than the volumes required to carry out the necessary measurements in order to limit the level of theoretical systematics \citep{desiCosmoSim2018}. 

Ideally, these simulations would solve equations of the physics of baryons and dark matter across cosmic time. Hydrodynamic simulations that account for the intricate physics that drives the formation of galaxies, however, are computationally expensive. The cost of simulating detailed physics that accounts for baryons in a volume that cosmological surveys require renders such simulations infeasible. Currently available hydrodynamical simulations, e.g. EAGLE \citep{Crain2015}, IllustrisTNG  \citep[presented in][and others]{Naiman2018, Nelson2018, Pillepich2018}, and Massive Black II \citep{Khandai2015}, cover volumes that are much smaller than what is required for cosmological surveys' needs.

While insufficient in volume, hydrodynamic simulations offer the potential for direct simulation of physical details behind galaxy formation and evolution. This property makes this class of simulations useful for  informing the methods that produce realistic galaxy populations more quickly and at lower computational cost. 

One way to circumvent the computational expense of running a full hydrodynamic cosmological simulation is to consider a dark matter-only N-body simulation, in which the equations of gravity only are solved, substantially bringing down computational costs. The simulation is then ``populated'' with galaxies following some algorithm, resulting in a catalogue of galaxies with properties and distribution that should be expected in a universe like the one that the N-body simulation represents. Methods for populating N-body simulations with galaxies are able to produce the cosmological-scale mock data that modern surveys require.

These methods can be broadly classified as physical, statistical, and statistical-empirical. The physical approach encompasses semi-analytic models (SAMs) \citep[e.g.][]{White1991, Kauffmann1993, Cole1994, Somerville1999, Cole2000, Baugh2006, Gonzalez-Perez2014, Croton2016, Lacey2016, Baugh2019}.  Statistical methods include biased dark matter \citep[e.g.][]{Cole1998, White2014}, halo occupation distributions (HOD) \citep[e.g.][]{Benson2000, Peacock2000, Berlind2002, Berlind2003}, and conditional luminosity functions \citep[e.g.][]{Yang2003, Cooray2006, Yang2008}. Statistical-empirical approaches include subhalo abundance matching (SHAM) \citep[e.g.][]{Vale2004, Kravtsov2004, Conroy2006} and its modifications \citep[e.g.][]{Skibba2009, Guo2016}. 

SHAM is a method of populating dark matter subhaloes with galaxies by matching the cumulative abundance functions of a dark matter halo property (commonly, subhalo dark matter circular velocity or mass) to the luminosity function or a similar cumulative distribution function of a galactic property. A variety of works have proposed that circular velocity, v\textsubscript{circ}, measured at various times in a subhalo's lifetime, may be an appropriate connector of host subhaloes to galaxies \citep[e.g.][]{Conroy2006, Masaki2013b, Reddick2013, Chaves-Montero2016}. 

A number of approaches adding scatter to a SHAM mock have been proposed, such as sampling a probability distribution \citep{Guo2016, Chaves-Montero2016}, fitting a parametrized model to a hydrodynamic simulation and sampling the resulting likelihood \citep{Chaves-Montero2016}, adding scatter to SHAM-style assignment of galaxy colours \citep{Yamamoto2015, Masaki2013b}, deconvolution \citep{Reddick2013} and shuffling with a fixed scattering magnitude, used in \citet{Mccullagh2017}, as well as the method described in this work.

SHAM offers the advantage of using a cosmological model's predictive power for the number and properties of subhaloes, as well as their relation to their host haloes while requiring few, if any, parameters \citep{Reddick2013}. Cosmological simulations that resolve subhaloes alleviate the need for assumptions about the occupation number and distribution of halo substructures, which are necessary for statistical  models, such as HODs. Implementations of SHAM have been shown to reproduce observed quantities that include the two-point correlation function \citep[e.g.][]{Conroy2006, Reddick2013, Lehmann2017}, three-point statistics \citep[e.g.][]{Tasitsiomi2004, Marin2008}, galaxy-galaxy lensing \citep[e.g.][]{Tasitsiomi2004}, and the Tully-Fisher relation \citep[e.g.][]{Desmond2015}.

The ultimate goal of this research is to produce a mock galaxy catalogue that closely mimics data that will be observed in DESI's Bright Galaxy Survey  \citep{DESI2016}. The Rosella mock catalogue described here uses SHAM to populate the P-Millennium N-body simulation (described in Section~\ref{sec:PMill}) with galaxies. Our approach provides rest-frame $r$-band absolute magnitudes and \gr\footnote{We denote absolute magnitudes and colours k-corrected to redshift 0.1 with the superscript \textsuperscript{0.1}.} \ colours assigned with algorithms described in sections~\ref{sec: luminosity algorithm} and~\ref{sec: colour assignment}, as well as positions, velocities, and host dark matter subhalo masses from P-Millennium. This work creates absolute magnitudes and colours k-corrected to \zapp 0.1 because this is the redshift used in papers that form the basis of our mock, for instance \citet{Zehavi2011} and \citet{Smith2017}. 

Our method for galaxy colour and luminosity assignment offers novel developments, namely the magnitude depth, volume, scatter, and the inclusion of subhalo history information.
Rosella's method of including scatter in the luminosity data uniquely conserves the target luminosity function, which enables the assignment of galaxy luminosities as faint as \Mhr $\sim -17.5$. We discuss this property in Section~\ref{sec:scatter}.

While we focused the detailed tuning of the mock presented in this paper on the needs of the DESI Bright Galaxy Survey (BGS), the mock can be used for other low-redshift galaxy surveys that might benefit from a \zapp 0.2 reference mock (e.g. the WAVES\footnote{\url{https://wavesurvey.org/}} survey in 4MOST). Furthermore, the method behind Rosella can be used to create galaxy mocks at other redshifts and, with some additional steps, extended into a lightcone mock. The method can thus benefit any survey that probes volumes similar to those covered by Rosella (see Section~\ref{sec:PMill} for details).

We have chosen to create this implementation of Rosella at the simulation snapshot that corresponds to redshift 0.203. The choice is motivated by the needs of the BGS. BGS will take the spectra of relatively bright galaxies during bright observing time. Consequently, its selection of target galaxies places the median redshift for future BGS observations at \zapp 0.2. Rosella will be useful as a reference mock for BGS, for fulfilling tasks that include analysing survey biases and calibrating approximate mocks that meet the volume and abundance requirements of the experiment \citep{desiCosmoSim2018}.

We evaluate the closeness of the match between our mock and real data by comparing the luminosity- and colour-dependent clustering of our mock's galaxies against previously published clustering of similar galaxy populations in existing observational and mock data.

This paper is organised as follows. Section~\ref{sec:methods} describes the N-body simulation that Rosella is built on and outlines the methodology behind our work. Section~\ref{sec:properties} described the properties of the Rosella mock, including the luminosity function, the luminosity- and colour-dependent clustering of the galaxies in Rosella, and the colour bimodality of galaxies in Rosella. Section~\ref{sec:conclusion} presents our main conclusions.

Throughout this work, $r$-band absolute magnitudes and $(g-r)$ colours are given in AB magnitudes, as defined for the SDSS system \citep[e.g.][]{Blanton2003}.

\section{SHAM with P-Millennium for the DESI Bright Galaxy Survey}\label{SHAM}\label{sec:methods}

A mock catalogue tailored for the needs of BGS already exists: it is a lightcone mock constructed with an application of HOD to the Millennium-XXL (MXXL) simulation \citep{Smith2017}. 
However, that mock catalogue has some limitations. The catalogue described in this paper can address these limitations. The simulation we use here, P-Millennium, offers high mass resolution that enables the tracking of fainter galaxies and the creation of a mock catalogue tailored with the scientific requirements of DESI's Bright Galaxy Survey (BGS) in mind.

\subsection{Simulation: P-Millennium}\label{sec:PMill}

The Planck Millennium N-body simulation (hereafter P-Millennium) is a high-resolution dark matter-only simulation of a 800 Mpc periodic box \citep{Baugh2019}. It is part of the `Millennium' series \citep{Springel2005Nature, Boylan-Kolchin2009MNRAS} of dark matter-only simulations of large-scale structure formation in cosmologically representative volumes carried out by the Virgo Consortium\footnote{\url{http://virgo.dur.ac.uk/}}.

P-Millennium is run using cosmological parameters given by the best-fit $\Lambda$ cold dark matter ($\Lambda$CDM) model to the first-year Planck cosmic microwave background data and measurements of large-scale structure in the spatial distribution of galaxies \citep{Planck2014}. The analysis of the final Planck dataset has introduced little change to these cosmological parameters \citep{Planck2018}. See Table~\ref{table: Pmill parameters} for a summary of the specifications of the P-Millennium run. 

\begin{table}
\resizebox{0.94\columnwidth}{!}
{
\begin{minipage}{\columnwidth}
\caption[Selected cosmological parameters of the P-Millennium simulation.]{Selected cosmological parameters of the P-Millennium simulation:  
 (1) $\Omega_{\textup{M}}$, present-day matter density in units of the critical energy density of the Universe,
 (2) $\Omega_{\textup{b}}$, the baryon density parameter, 
 (3) $\Omega_\Lambda$, the energy density parameter of the cosmological constant, $\Lambda$,
 (4) n\textsubscript{spec}, the spectral index of the primordial density fluctuations,
 (5) $h$, the reduced Hubble parameter, $h = \textup{H}_0$ / (100 km s\textsuperscript{-1} Mpc\textsuperscript{-1}),
 (6) $\sigma_8$, the normalisation of the density fluctuations at the present day, 
 (7) N\textsubscript{p}, the number of particles,
 (8) L\textsubscript{box}, the simulation box length,
 (9) M\textsubscript{p}, the mass of individual particles in the simulation, and 
 (10) M\textsubscript{h}, the minimum mass of a resolved halo, corresponding to 20
particles.  See \citet{Baugh2019} for further details.}
\begin{center}

\begin{tabular}{| c | c |}
	\hline
	Parameter name & Value in P-Millennium \\ \hline
	$\Omega_\textup{M}$ & 0.307 \\
	$\Omega_\textup{b}$  &  0.0483 \\
	$\Omega_\Lambda$  & 0.693 \\
	$n_{\textup{spec}}$  & 0.9611 \\
	$h$ & 0.6777 \\
	$\sigma_8$ & 0.8288 \\
	N$_{\textup{p}}$ & 5040$^3$ \\
	L$_{\textup{box}}\ [h^{-1}$ Mpc] & 542.16\\
	M$_{\textup{p}}\ [h^{-1}\ \textup{M}_\odot]$ & $1.06\ \times\ 10^8$ \\
	M$_{\textup{h}}\ [h^{-1}\ \textup{M}_\odot]$ & $2.12\ \times\ 10^9$ \\
	\hline
\end{tabular}

\label{table: Pmill parameters}
\end{center}
\end{minipage} }
\end{table}

The mass resolution of P-Millennium is $1.06\ \times\ 10^8\ \ \textup{M}_\odot \ h^{-1}$ per particle, with $5040^3$ particles representing the matter distribution \citep[for a detailed comparison to other simulations in the Millennium suite, see][]{Baugh2019}. The lowest resolved halo mass in P-Millennium is $2.12\ \times\ 10^9\ \textup{M}_\odot \ h^{-1}$.  This makes the simulation appropriate for SHAM, since the simulation's mass resolution lets SUBFIND \citep{Springel2001} resolve dark matter halo substructures, subhaloes --- a central component for creating a mock using SHAM (see Section~\ref{sec: SHAM with PMill} for a discussion). 

The low halo mass limit in P-Millennium allows us to create a mock with a faint absolute magnitude limit that reaches beyond the minimum luminosity cutoffs offered in other mock catalogues. For example, the \textit{Buzzard} catalogue, presented in \citet{DeRose2019}, creates a reference mock that models the galaxy distribution down to roughly\footnote{We define $r$-band absolute magnitude dependent on \textit{h} and \\with boundaries defined at \zapp0.1  as  \Mhr $\equiv$ \Mhrl , where $h$ is the dimensionless constant given as H\textsubscript{0} = 100 $h$ km s$^{-1}$ Mpc$^{-1}$.} \Mhr  $=-18.2$, saying that ``the SHAM catalog is not strictly complete'' down to that absolute magnitude. As discussed in Section~\ref{sec: LF}, Rosella can be fully complete for galaxies as faint as \Mhr $=-17.5$, depending on the choice of scatter that is implemented.

\subsubsection{Tracing P-Millennium subhalo histories}\label{sec: subhalo histories}

\vpeak\  is the central quantity that allows us to connect dark matter subhaloes in our N-body simulation to the galaxies in our mock catalogue. Its definition is built on the quantity v\textsubscript{circ}, defined as:
\begin{equation}\label{eqn: vcirc}
\mathrm{v}_{\textup{circ}}(r, z) = \sqrt{\frac{\textup{GM(}z, <r)}{r}}
\end{equation}
\textit{r} here is the physical distance between the particle and the centre of the subhalo, \textit{z} is redshift, G is the gravitational constant, and M(\textit{z, <r}) is the mass enclosed within the radius \textit{r}, at redshift \textit{z}. Maximum circular velocity \vmax \ is the value of v\textsubscript{circ} at the radius at which v\textsubscript{circ} reaches its maximum:
\begin{equation}
\mathrm{v}_{\textup{max}}(z) = \textup{max}[\mathrm{v}_{\textup{circ}}(r, z)]
\end{equation}
\vpeak \ is the highest \vmax\ that a subhalo reaches over the course of its existence in the simulation:
\begin{equation}\label{eqn:vpeak}
\mathrm{v}_{\textup{peak}} = \textup{max}[\mathrm{v}_{\textup{max}}(z)]
\end{equation}

\begin{figure}
\centering
\includegraphics[width=\columnwidth]{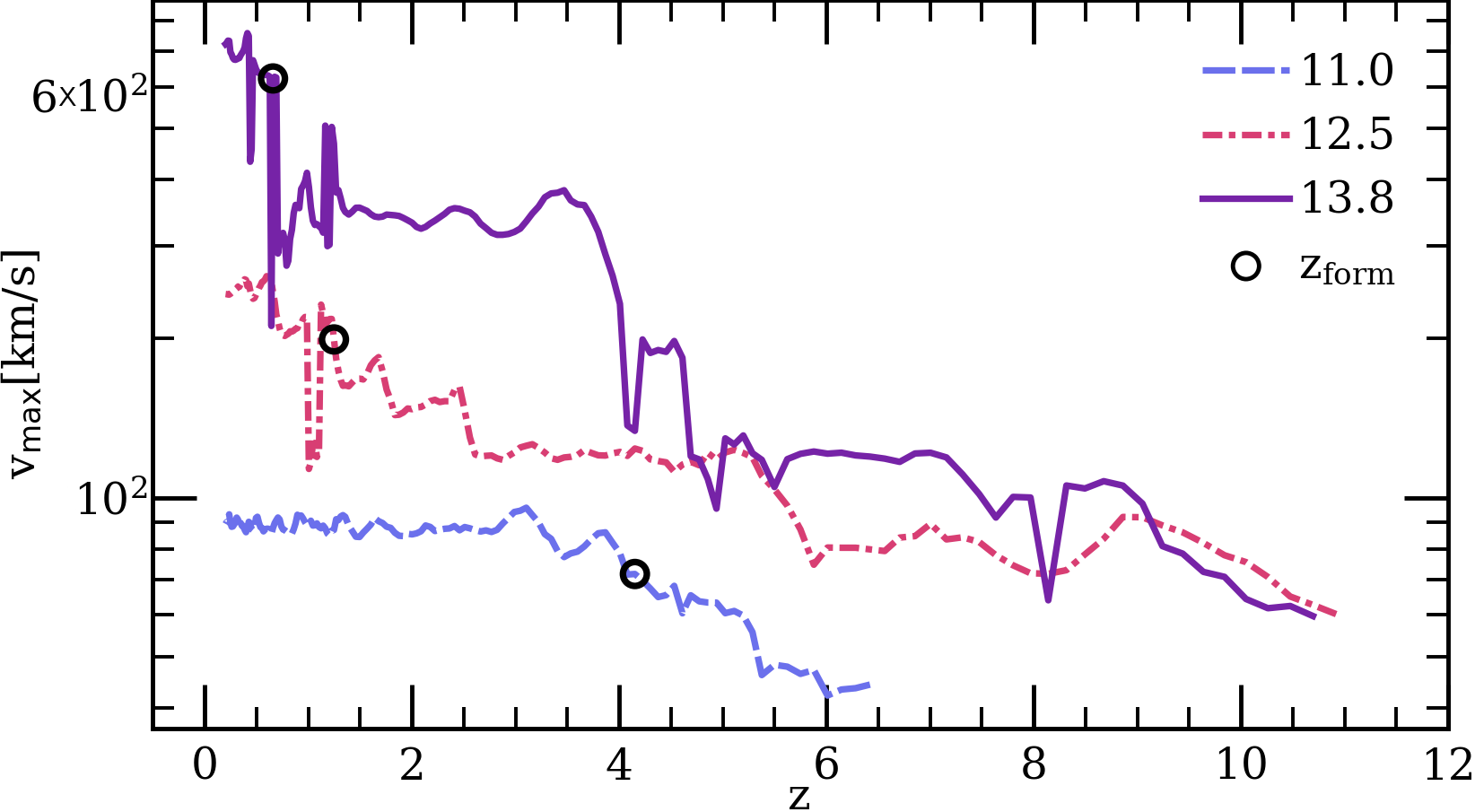}
\caption{Examples of maximum circular velocities as a function of redshift. The vertical axis shows the \vmax \ values of individual subhaloes at given redshifts $z$. Each line tracks the \vmax \ history of a subhalo that has mass M at \zapp 0.2, expressed as $\log_{10}[{M}/({M_\odot}\ h^{-1})]$, indicated in the legend. Black circles indicate \zform \ values for these subhaloes, calculated using the method described in Section~\ref{sec: how we calculate zform}.
}
\label{fig:vmax histories}
\end{figure}

To calculate \vpeak, as well as a proxy for a subhalo's age, \zform, which we describe in Section~\ref{sec: how we calculate zform}, we compile the histories of \vmax \ values that individual P-Millennium subhaloes reach over the course of the simulation. This non-trivial operation is described and discussed in greater detail in \citet{Safonova2019}\footnote{The code for this procedure, along with the code used to complete the rest of the Rosella methodology, is stored in a private repository at \url{https://github.com/safonova/pmillennium-sham}.}. Examples of such histories are plotted in Fig.~\ref{fig:vmax histories}. We generated a full dictionary of subhalo histories for subhaloes found at the P-Millennium snapshot corresponding to $z=0.203$. The histories show transitory features that appear like short-lived drops in \vmax  , perhaps related to mergers and the difficulties of tracking subhaloes during mergers \citep[e.g.][]{Behroozi2012}. In Section~\ref{sec: how we calculate zform}, our definition of \zform \ makes these features inconsequential.

\subsubsection{Definition of formation redshift}\label{sec: how we calculate zform}

\begin{figure}
	\includegraphics[width=\columnwidth]{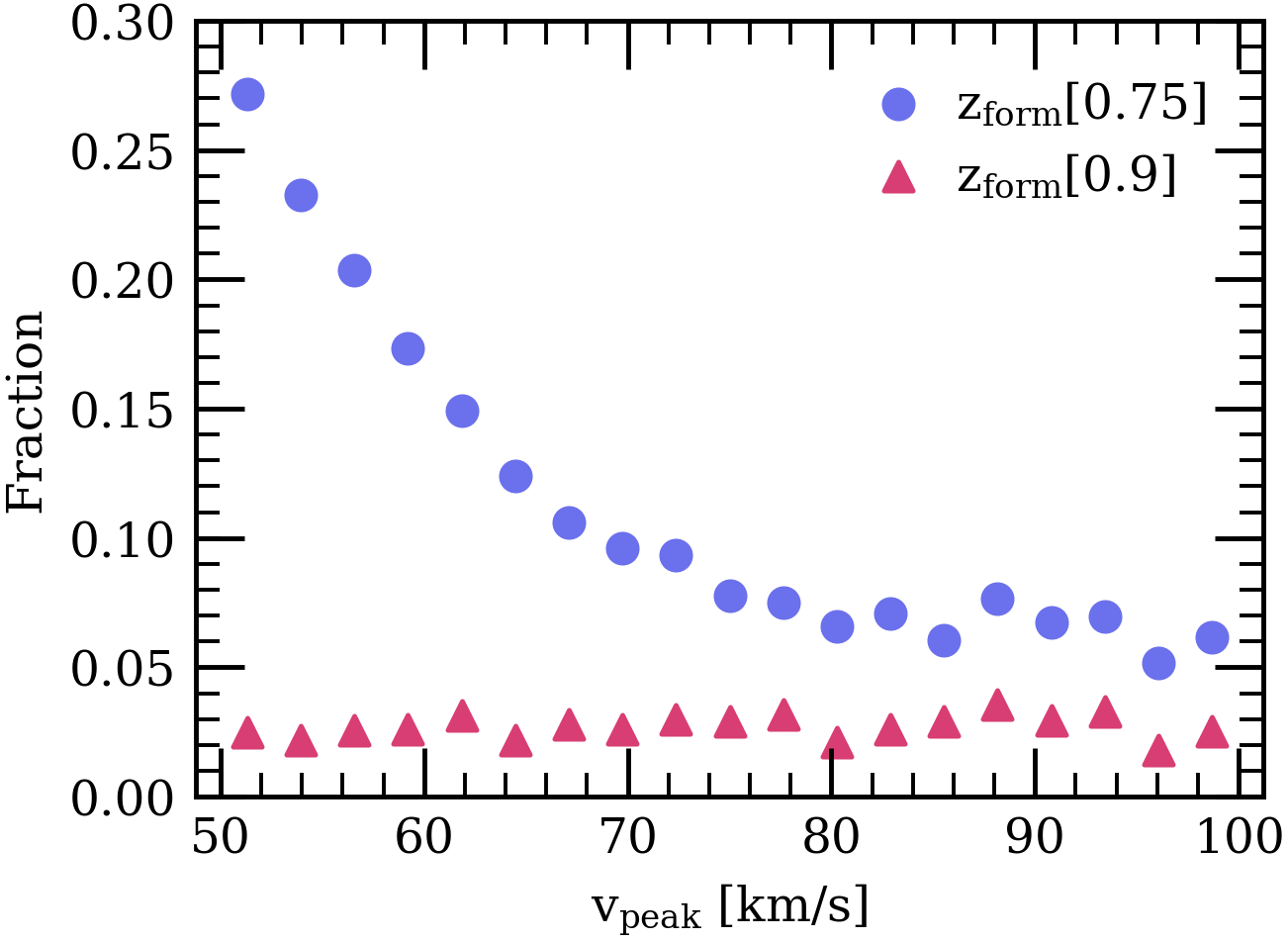}
    \caption{Fraction of subhaloes with subhaloes for which we only have lower limits on the value of \zform , plotted as a function of \vpeak. The value of \zform \ becomes important in our colour assignment scheme. This plot informs us about the completeness of the mock's colour assignments. Subhaloes for which only a lower limit on \zform \ can be set are those whose earliest identified progenitor has \vmax \ > \vform. The blue circles and red triangles correspond respectively to using 
    $f=0.75$  and  $f=0.9$ in equation~\ref{eqn: vform} that defines \vform. }
    \label{fig:colour resolution}
\end{figure}

In order to assign colours to Rosella galaxies, we compute each subhalo's ``formation redshift'', \zform , which serves as a proxy for a subhalo's age. The choice to connect galaxy colours to the ages of their host subhaloes stems from the idea that older subhaloes are likely to have older and, consequently, redder stellar populations \citep[e.g.][]{Mo2010, Hearin2015}. We compute an individual subhalo's \zform \ based on the criterion that \zform \ corresponds to the maximum output redshift at which a subhalo's \vmax \ exceeds v\textsubscript{form}:
\begin{equation}\label{eqn: vform}
\textup{v}_{\textup{form}} = f\, \textup{v}_{\textup{peak}},
\end{equation}

Here, $f$ is a free parameter.
We identify the two output redshifts between which \vform \ is located and interpolate between them to get \zform. If the pre-\vpeak \ history of the subhalo consists only of \vmax \ values greater than v\textsubscript{form}, \zform \ is set to the redshift corresponding to the earliest snapshot at which the subhalo is found.

It is possible to adjust the $f$ parameter, or even the relationship between v\textsubscript{peak} and v\textsubscript{form}, to tune the mock data produced with the model presented here. We have considered two values of $f$, 0.75 and 0.9. We have noted that $f=0.75$ produces a more favourable match to clustering data (see Section~\ref{sec: colour-dependent clustering}). 
Expression~\ref{eqn: vform} is inspired by the works of \citet{Masaki2013a} and \citet{Yamamoto2015}; however, those papers work with \vmax \ instead of \vpeak . Nonetheless, the v\textsubscript{form} in \citet{Masaki2013a} and \citet{Yamamoto2015} has a similar underlying structure to the criterion that serves as a proxy for subhalo age in our methodology.

Choosing lower values of $f$ shifts the distribution of \zform \ to higher redshift. This can be problematic if the true value of \zform \ is then larger than the redshift, $z$\textsubscript{max},  of the first P-Millennium snapshot where a subhalo is found. Nevertheless, to create informative galaxy colours based on subhalo \vmax \ history, the lowest possible \zform \ values provide the most robust information about subhalo history. Thus, we conduct a test of the strength of subhalo history information with  $f=0.75$ and $f=0.9$.  Fig.~\ref{fig:colour resolution} shows this test: the fraction of subhaloes for which \zform \ is assigned as the highest $z$ at which the subhalo is found in P-Millennium in bins of \vpeak \ at \zapp 0.2. We consider this condition to describe a ``poorly defined \zform'', as it does not include information about the subhalo's history when its mass lies below the P-Millennium halo mass resolution. Thus, the blue circles in Fig.~\ref{fig:colour resolution} trace the fraction of subhaloes that have reached a \vpeak \ given on the horizontal axis by \zapp 0.2 but have a poorly defined \zform.

 Fig.~\ref{fig:colour resolution} illustrates the impact of the P-Millennium resolution on the choice of $f$: the smaller $f$ is, the larger the limit on \vpeak\ has to be to ensure that subhalo progenitor trees are sufficiently complete. Typically with $f=0.75$, we can consider P-Millennium to be complete for subhaloes with \vpeak $\ \geq$ 75 km/s. We discuss this further in Section~\ref{sec: LF}.

\subsubsection{Definition of central galaxies}\label{sec: centrals definition}

In Rosella, every central galaxy is located at the position of the most gravitationally bound particle in its host friends-of-friends halo. Galaxies in subhaloes outside the central gravitational well of a friends-of-friends halo are considered satellites.

\subsection{SHAM with P-Millennium}\label{sec: SHAM with PMill}
There are several advantages to SHAM as a method for populating P-Millennium with galaxies.

Implementing SHAM is relatively quick compared to a physical method, such as a full semi-analytic galaxy formation model. Additionally, it can be arbitrarily tuned to reproduce certain statistics, as it includes empirical components in its methodology through its free parametrization via both functional models and numerical values.

SHAM is ideal for the analysis of groups and clusters for which BGS data may be used in the future and for which HOD models might not be complex enough. For example, it is not clear whether the mitigation techniques planned for DESI can recover statistics affected by assembly bias. BGS will benefit from mock data that includes halo assembly bias.

\subsubsection{Assembly bias with SHAM}\label{sec:assembly bias}
Halo assembly bias\ describes the phenomenon that dark matter halo clustering depends on properties besides halo mass, including but not limited to formation time, concentration and spin \citep[e.g.][]{Gao2005, Wechsler2006, Gao2007}. For a given halo mass, clustering is stronger in dark matter haloes that form at earlier times. The dependence of clustering on halo formation time increases with decreasing halo mass \citep{Gao2005}. 

\vmax \ characterises the depth of gravitational potential. At fixed halo mass, v\textsubscript{max} is directly related to halo concentration \citep[e.g.][]{Conroy2006, Zehavi2019}. As halo concentration has been suggested to be a quantity that can track galaxy assembly bias, it offers the possibility of lifting the systematic effects of galaxy assembly bias in mock data. However, \vmax \ describes the present state of a subhalo, which may miss some of the historical information contained in, for example, \vpeak. \citet{Chaves-Montero2016} offers one comparison of the qualities that v\textsubscript{circ}-related SHAM proxies impart on mock data.

This presents a problem for halo occupation models that assume the independence of the distribution and properties of galaxies from their environment beyond halo mass \citep{Gao2005}. Abundance matching on subhalo quantities that include information about their history, such as peak circular velocity \vpeak\  or satellite subhalo accretion mass M\textsubscript{acc}, may lift part of this assumption of distribution-environment independence in the galaxy-halo occupation relation. 

By incorporating a proxy that implicitly accounts for subhalo assembly history, \vpeak , a SHAM catalogue can be more informative when investigating the effects of assembly bias on observational data and computing statistics that may be affected by it, compared to a traditional HOD mock. Some work, however, has been done that allows tunable assembly bias to be included in modified HOD methods \citep[e.g.][]{Hearin2016} and SHAM methods \citep[e.g.][]{Contreras2020}.
\subsubsection{Algorithm for luminosity assignment}\label{sec: luminosity algorithm}
\begin{figure*}
\centering
\includegraphics[width=2\columnwidth]{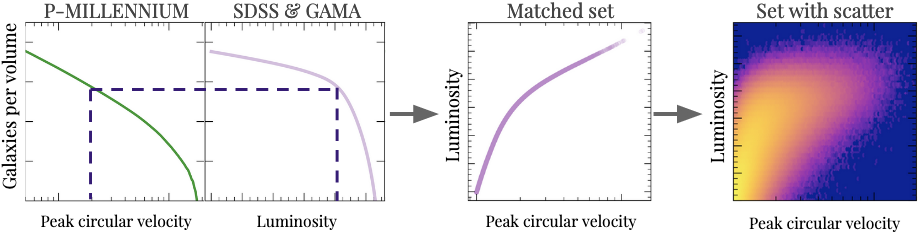}
\centering
\caption[Illustration of SHAM]{Illustration of assigning luminosities to galaxies using SHAM with the addition of scatter. The first two panels show abundance relations: left panel shows the abundance of subhaloes as a function of their \vpeak \ in our N-body simulation, and the second panel from the left shows the luminosity function. For a given subhalo with a known \vpeak, we follow the dashed line to match its abundance in a simulation to a luminosity value that has the same abundance in observations. Repeating this matching for a set of subhaloes produces a set of points that form a tight line, as seen on the third panel from the left. Beginning with the third panel, we add scatter to the luminosity-\vpeak \ data set. The data with the added scatter no longer follows a line in luminosity-\vpeak \ space. The rightmost panel shows the logarithmic density of this illustrative set of data points in luminosity-\vpeak \ space after the addition of scatter. The scatter method used here preserves the luminosity function of the no-scatter counterpart of this SHAM data set, as discussed in Section~\ref{sec: LF}.}
\label{fig:scatter SHAM}
\end{figure*}

We assign luminosity values to galaxies in our mock catalogue by assuming that a galaxy occupies every dark matter subhalo that satisfies a minimum value of \vpeak . We assume that galaxy luminosities correlate with \vpeak .

\vpeak , by construction, includes information about a subhalo's formation history. When we populate satellite subhaloes with galaxies, \vpeak \ allows us to account for the historical values of that subhalo's \vmax , thus mitigating the influence of effects like dark matter mass stripping as a consequence of mergers. There has been evidence of subhaloes with higher \vpeak \ values tending to have higher concentration  and earlier formation times, which are some of the properties associated with assembly bias \citep[see][and reference therein]{Xu2018}. \vpeak \ can have some downsides, such as some post-merger transient features which may not correlate with changes in galaxy properties \citep{Chaves-Montero2016}.

Initially, we assume that subhalo \vpeak\  follows a monotonic relation with the galaxy absolute magnitude in the \textit{r} band, \Mhr. In the first step of luminosity assignment, we operate under the assumption that the relation between magnitude and \vpeak\  are one-to-one, but that assumption is no longer applicable once we add scatter to the mock data. For the first, no-scatter, stage of our algorithm, the assumed relation between \Mhr \ and \vpeak \ can be expressed as:
\begin{equation}
		\textup{n}_{\textup{g}}(< M_r^h) = \textup{n}_{\textup{h}}(> \textup{v}_{\textup{peak}})
\label{eqn: no scatter SHAM}
\end{equation}
Here, n\textsubscript{g} is the number density of galaxies of a given \Mhr \ or brighter, and n\textsubscript{h} is the number density of subhaloes of a given \vpeak \ or higher. In other words, the magnitude \Mhr \ that we assign to a galaxy in a subhalo with \vpeak \textsubscript{,i} is set by matching the abundance of subhaloes with \vpeak >\vpeak \textsubscript{,i} to the abundance of galaxies with \Mhr <\Mhr \textsubscript{,i}.

We follow a number of specific steps to assign magnitude values to the galaxies in our sample:
\begin{enumerate}
		\item Get the evolving $r$-band galaxy luminosity function (LF) using the SDSS $r$-band LF \citep{Blanton2003} and the GAMA $r$-band LF \citep{Loveday2012}. The combined smooth LF used here is the one that \citet{Smith2017} used for the development of a DESI BGS lightcone mock catalogue. We call this set of reference data the `target luminosity function', as it is the LF that we aim to replicate in our mock.
		\item Perform SHAM with zero scatter using the target luminosity function with the monotonic relation between luminosity and \vpeak \ in equation~\ref{eqn: no scatter SHAM}.  \citet{Chaves-Montero2016} and \citet{Mccullagh2017} also used this relation as the basis of their SHAM assignments. Fig.~\ref{fig:scatter SHAM} offers an illustration of the process.
		\item Add luminosity-dependent scatter, following \cite{Mccullagh2017}\footnote{This method effectively shuffles the ranks while maintaining the originally assigned set of luminosities. Hence, it does not perturb the cumulative luminosity function, and no deconvolution is necessary, unlike other methods of adding scatter to SHAM data.}, using a magnitude-dependent scatter $\sigma(M_r^h)$ to produce results that are illustrated in Fig.~\ref{fig:scatter SHAM}. See below for more details on the scatter algorithm.
\end{enumerate}

Before scatter, we use the galaxy cumulative luminosity function (LF) down to \Mhr $=-10$ for the purposes of fully utilizing our LF-preserving scatter method. After scatter, we keep galaxies that are \Mhr$=-17.5$ or brighter, which corresponds to a minimum \vpeak \ of $\sim$ 75 km/s. Our choice to limit the analysis to subhaloes with \vpeak \ $\geq 75$ km/s is motivated by the P-Millennium resolution (see Section~\ref{sec: how we calculate zform} and Section~\ref{sec: LF}).
\subsubsection*{Adding scatter to SHAM}\label{sec:scatter}

 The approach described here uses a magnitude-dependent scatter magnitude $\sigma(M_r^h)$ \citep[called $\Delta M_r^h$ in][]{Mccullagh2017} to produce results that are illustrated in Fig.~\ref{fig:scatter SHAM}. We execute the following four steps to add luminosity-dependent scatter to the magnitude values of the galaxies in our sample:
\begin{enumerate}\label{list:scatter steps}
    \item  Assign a magnitude without scatter, \Mhr, to every galaxy using the method described above;
    \item For every galaxy, draw a new magnitude, $M_r^{h\prime}$, from a Gaussian distribution clipped at 2.5 $\sigma(M_r^h)$, with the mean equal to the galaxy's \Mhr \ value and the standard deviation $\sigma(M_r^h)$ computed as a function of the galaxy's absolute magnitude. In this work, $\sigma(M_r^h)$ is given by a smooth step function of the form
	\begin{equation}\label{eq:deltamr}
		 \sigma(\textup{M}_r^h) = \alpha + \beta \tanh(\textup{M}_r^h-M^h_{r, \textup{ref}})
	\end{equation}
where $\alpha$, $\beta$ and $M^h_{r, \textup{ref}}$ are free parameters that we can tune to match clustering (Section~\ref{sec:tuning} discusses tuning the parameters in this method). A variable, luminosity-dependent $\sigma(\textup{M}_r^h)$ was chosen to create luminosity-threshold clustering of the data that matches the clustering of galaxies in observations and existing mock data, (for the clustering analysis, see Section~\ref{sec:luminosity clustering}). To create the Rosella catalogue presented here, we use the following parameter values for the model in equation~\ref{eq:deltamr}:
			\subitem $\alpha = 0.8$; $\beta = 0.4$; $M^h_{r, \textup{ref}} = -20$
	\item Rank galaxies in order of the new magnitude, $M_r^{h\prime}$;
	\item Rank subhaloes in order of their \vpeak\  values;
	\item Place galaxies in subhaloes so that the galaxies with the brightest $M_r^{h\prime}$ are located in the subhaloes with the largest \vpeak \ values;
	\item Assign each galaxy's original magnitude, \Mhr , to the galaxy's final location computed in the step above.
\end{enumerate}

\subsubsection{Luminosity-dependent colour assignment}\label{sec: colour assignment}\label{sec:vpeak-dep zform cdf}

A number of methods that have built upon original abundance matching assign colours to galaxies in gravity-only simulations based on (sub-)halo age or environment \citep[e.g.][]{HearinWatson2013, Masaki2013a, Hearin2014, Yamamoto2015}.

A common approach to assigning galaxy colours in a SHAM-like paradigm matches subhaloes' directly simulated (sub-)halo property, such as v\textsubscript{max} or \vpeak, and a secondary (sub-)halo property that serves as a proxy for its age \citep[see][]{Masaki2013a, KulierOstriker2015, Yamamoto2015}. This is the so-called ``age model''  of the dark matter halo-based prediction of galaxy colour. The approach is based on the notion that older galaxies should contain older, and, consequently, redder, stellar populations. Thus, if galaxy colour can be used as a measure of stellar population age when we analyse observations, we should be able to reverse the process and assign colours to simulated galaxies based on the ages of their subhaloes.

\begin{figure}
\centering
\includegraphics[width=\columnwidth]{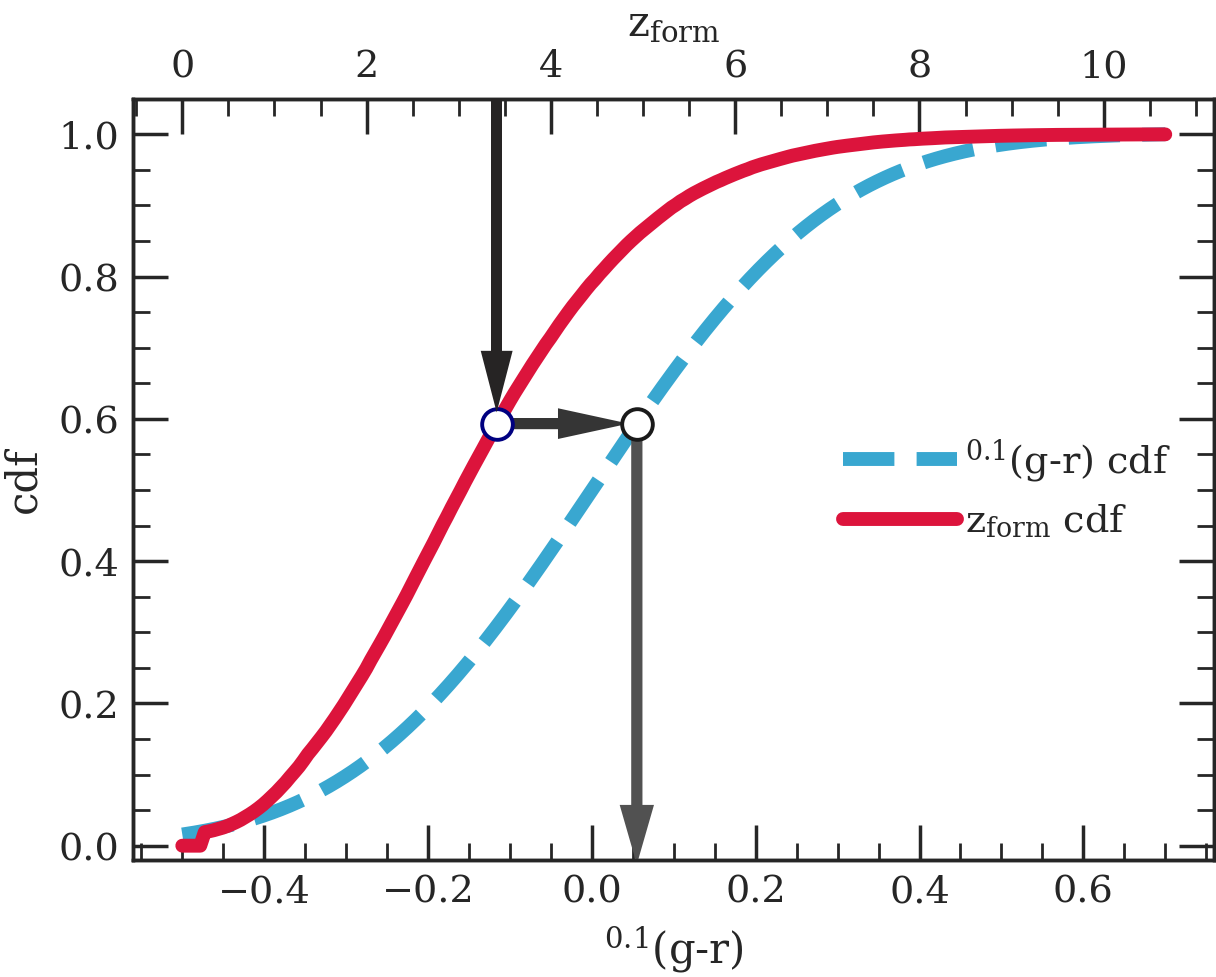}
\caption[Illustration of colour assignment for a single subhalo]{Illustration of colour assignment for a single subhalo. The top horizontal axis shows formation redshift, \zform ; the bottom horizontal axis shows \gr . The vertical axis shows cumulative distribution function (cdf) values for the red (solid) and blue (dashed) curves. The red (solid) curve is the cdf of subhalo \zform \ values for a given \vpeak . 0 on this curve means that no subhaloes of the given \vpeak \  should be expected to have that or lower \zform . 1 on the red (solid) curve signifies that all subhaloes of the given \vpeak \ should be expected to have lower \zform \ values. The red (solid) curve is computed by interpolating between \zform \ cdf curves calculated in bins of \vpeak \ (see the left panel of Fig.~\ref{fig: zform and color cdfs} for examples). The blue (dashed) curve is given by equation~\ref{eqn: colour cdf} and is the cdf of \gr \ for this subhalo's \Mhr . }
\label{fig:single subhalo colour assignment}
\end{figure}

The procedure for the assignment of  \gr \ colours to Rosella galaxies comprises three steps, illustrated in Fig.~\ref{fig:single subhalo colour assignment}, and is built around two notions. Galaxy colour bimodality analyses \citep[e.g.][]{Baldry2004a} show that brighter galaxies tend to be redder across both blue and red populations of galaxies. Thus, we begin colour assignment by calculating a cumulative distribution function of \gr , conditional on \Mhr , individually for each galaxy. We describe this procedure in Section~\ref{sec:lum-dep gr cdf}.

\begin{figure*}
\centering
\includegraphics[width=2.1\columnwidth]{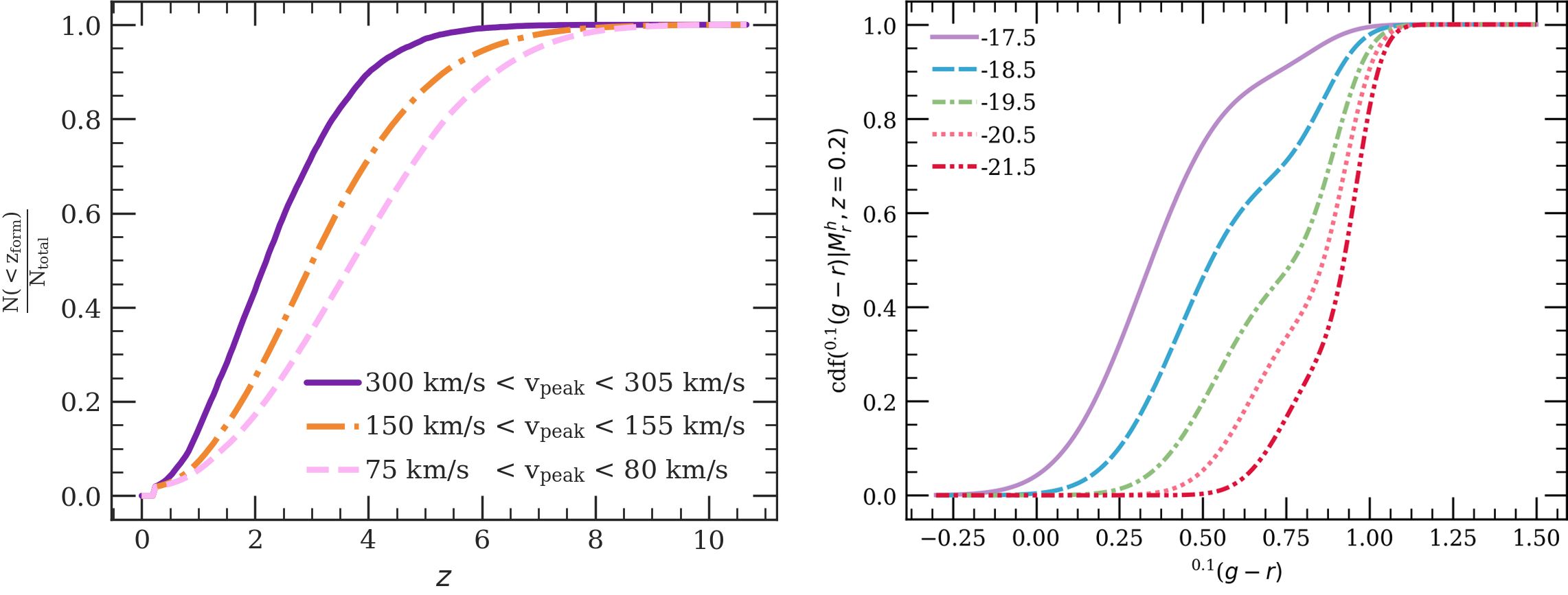}
\caption{Left panel: Example cumulative distribution functions of formation redshift in bins of \vpeak. The vertical axis displays the fraction of subhaloes with a given formation redshift \zform \ or lower. The horizontal axis is the formation redshift \zform. Curves are colour coded by bins of \vpeak. Right panel: Cumulative distribution functions of \gr \ at $z=0.2$ for a selection of \Mhr \ values, as indicated in the legend. The functional form of these distributions is given in equation~\ref{eqn: colour cdf}.}
\label{fig: zform and color cdfs}
\end{figure*}

The second part of our colour assignment procedure builds upon the correlation between galaxy colour and age \citep[e.g.][]{Mo2010, Hearin2015, Chaves-Montero2020}. 
Our colour assignment method requires relating the cumulative distribution function (cdf) of \zform \ at a given \vpeak \ value to the cdf of \Mhr-dependent \gr \ colour distributions. However, we do not know the distribution of \zform \ for any individual galaxy with its unique \vpeak \ a priori. We therefore construct cdfs of \zform \ from subsample populations of galaxies in narrow bins of \vpeak . Examples of such \zform \ distribution functions are provided in the left panel of Fig.~\ref{fig: zform and color cdfs}.

Note that median \zform \ values decrease with increasing median \vpeak \ values in the left panel of Fig.~\ref{fig: zform and color cdfs}, which is a consequence of the hierarchical formation of structure in the Universe. During colour assignment, we interpolate between the full set of curves covering the full range of \vpeak \ values to find an appropriate \zform \ cdf for each subhalo.

As the final step in \gr \ assignment, we find each subhalo's position on the \vpeak -dependent distribution of \zform \ and translate it to a \gr \ value for the galaxy residing in it, as illustrated in Fig.~\ref{fig:single subhalo colour assignment} and discussed in more detail below.

\subsubsection{Luminosity-dependent galaxy colour distributions}\label{sec:lum-dep gr cdf}

The colour-magnitude diagram of observed galaxies has a bimodal distribution \citep[e.g.][]{Baldry2004a} that can be described as a sum of components that correspond to red and blue galaxy populations. To assign a colour to a galaxy in Rosella, we start with the empirical model for the observed bimodal luminosity-dependent distribution of \gr \ colours given in \cite{Smith2017}. 

At a given magnitude  \Mhr , it is assumed that the blue and red components of the \gr \ distribution functions each have Gaussian forms and that their combined cumulative distribution function (cdf) is given by
\begin{equation}\label{eqn: colour cdf}
\begin{split}
\textup{cdf}(M_r^h) = & f_{\textup{blue}}(M_r^h) \ \textup{G}(M_r^h)_{\textup{blue}} + \\
& (1-f_{\textup{blue}}(M_r^h)) \ \textup{G}(M_r^h)_{\textup{red}},
\end{split}
\end{equation}
where $f_{\textup{blue}}$ is the fraction of blue galaxies. This fraction is a function of
magnitude,
\begin{equation}\label{eqn: fblue}
f_{\textup{blue}} = 
    \begin{cases}
    0 & \mbox{if }  M_r^h < -26.571 \\
    0.46 + 0.07 (M_r^h + 20.) & \mbox{if } -26.571 \leq M_r^h < -19.539 \\
    0.4 + 0.2 (M_r^h + 20.) & \mbox{if } -19.539 \leq M_r^h > -17.173 \\
    \frac{1}{ 1 + \textup{exp}(-(M_r^h + 20.5))} & \mbox{if }  M_r^h > -17.173, \\
    \end{cases}
\end{equation}
while the mean and scatter of each of the Gaussian components are magnitude- and redshift-dependent, given in equation~\ref{eqn: g-r}. The sigmoid expression for the faintest galaxies ensures that the fraction of red galaxies slowly tapers off instead of meeting a sharp cutoff at a fixed magnitude, which makes our model slightly different from the prescription in \cite{Smith2017}.

We adopt relations from \cite{Smith2017} evaluated at $z=0.2$ as the mean and scatter of each of the Gaussian components in equation~\ref{eqn: colour cdf}\footnote{The $z$ term in the expression for $\textup{rms}(^{0.1}\textup{(}g-r\textup{)} |M^h_r)_{\textup{red}}$ contained a typo in Eq. 33 in \citet{Smith2017}. The correct formulation is \begin{equation}
\begin{aligned}
\operatorname{rms}\left(g-r \mid M_{r}\right)_{\operatorname{red}}(z) &=\operatorname{rms}\left(g-r \mid M_{r}\right)_{\operatorname{red}} \\
&+0.05(z-0.1)+0.1(z-0.1)^{2}
\end{aligned}
\end{equation}}:
\begin{equation}\label{eqn: g-r}
\begin{aligned}
\langle^{0.1}\textup{(}g-r\textup{)} |M^h_r\rangle_{\textup{blue}} & =  0.595 -0.11 (M^h_r + 20)\\
\textup{rms}(^{0.1}\textup{(}g-r\textup{)} |M^h_r)_{\textup{blue}} & =  0.14 + 0.02 (M^h_r + 20)\\
\langle^{0.1}\textup{(}g-r\textup{)} |M^h_r\rangle_{\textup{red}} & = 0.914 -0.032 (M^h_r + 20) \\
\textup{rms}(^{0.1}\textup{(}g-r\textup{)} |M^h_r)_{\textup{red}} & = 0.076 + 0.01(M^h_r + 20)
\end{aligned}
\end{equation}

The right panel in Fig.~\ref{fig: zform and color cdfs} shows examples of \gr \ colour cdfs for a selection of \Mhr \ values. 

We connect the cdfs of \zform \ to those of \gr , as illustrated in Fig.~\ref{fig:single subhalo colour assignment}, where the \zform \ cdf for a single subhalo is given by the red curve and is conditional on its \vpeak . Colour assignment consists of four steps:

\begin{enumerate}
    \item Compute the \gr\ cdf by applying a galaxy's \Mhr \ magnitude to equation~\ref{eqn: colour cdf};
    \item Compute the \zform \ cdf from the host subhalo's \vpeak \ value, as described in Section~\ref{sec:vpeak-dep zform cdf};
    \item Find the cdf value corresponding to the host subhalo's \zform , as shown by the top vertical arrow in Fig.~\ref{fig:single subhalo colour assignment};
    \item Determine the \gr \ value that matches the aforementioned cdf value, as demonstrated by the horizontal arrow in Fig.~\ref{fig:single subhalo colour assignment}. Assign this \gr \ value to the galaxy.
\end{enumerate}

\subsection{Tuning the catalogue}\label{sec:tuning}

The method has the following freedoms and free parameters: 

\begin{enumerate}
    \item The subhalo attribute connecting its present state to its history for age-matching colour assignment --- in the current method, this attribute is the distribution of \zform \ conditional on \vpeak \ and \Mhr \ (see Section~\ref{sec: how we calculate zform} and Fig.~\ref{fig:single subhalo colour assignment});
    \item  The functional form of $\sigma$(\Mhr ) in equation~\ref{eq:deltamr};
    \item The parameters $\alpha$, $\beta$, and \Mhr \textsubscript{ref} in equation~\ref{eq:deltamr};
    \item The specific definition of \zform \ (see Section~\ref{sec: how we calculate zform}), including $f$ in equation~\ref{eqn: vform}.
\end{enumerate}

By construction, our \zapp 0.2 mock is tuned to reproduce the galaxy LF, following the parametrization proposed by \cite{Smith2017}, which agrees with observational constraints provided by SDSS and GAMA. We also match, by construction, the luminosity-dependent colour distribution.

We tune the free parameters of the Rosella mock to match the observed luminosity- and colour-dependent clustering by comparing our results to the \citet{Smith2017} Millennium-XXL mock, as it represents observational data well. The Millennium-XXL mock fits the observational data at a range of redshifts,  but can be estimated at \zapp~0.2, the reference redshift of Rosella. In this work, we compare Rosella to the clustering of galaxies in the Millennium-XXL mock as it is presented in \citet{Smith2017}. The authors of \citet{Smith2017} use redshift ranges that correspond to SDSS volume-limited luminosity threshold samples in \citet{Zehavi2011}. We also compare our results to SDSS results presented in \citet{Zehavi2011}. For luminosity-dependent clustering, we examine redshift-space results in the context of existing mock data.

Full optimization of these parameters is beyond the scope of this paper, as that depends on the science goals for which Rosella and its methods are to be used.

To choose the value of $f$ in equation~\ref{eqn: vform}, we also consider the resolution of subhalo progenitors, as shown in Fig.~\ref{fig:colour resolution}. To choose the subhalo attribute for age-matching colour assignment, we also considered the galaxy colour bimodality discussed in Section~\ref{sec:colour bimodality in Rosella}.

We examined the effect of the choice of $f$ on the colour-dependent clustering of Rosella galaxies. The difference in the colour-dependent clustering between $f=0.9$ and $f=0.75$ (the default value) is small. Qualitatively, changing the value of $f$ from 0.75 to 0.9 slightly increases the gap between the clustering of red and blue galaxies. On small scales, 0.75 provides a better match, and we do not see the reason to increase the gap between red and blue galaxy clustering by setting $f$ to 0.9 on the larger scales.

The motivation for a nonzero $\beta$ in equation~\ref{eq:deltamr} is the observation that scatter driven by a constant $\sigma(M_r^h)$ produces unsatisfactory clustering results, generating a dataset with clustering that was consistently higher than that measured in observations, as shown in \cite{Zehavi2011}, particularly in the brightest samples. A luminosity-dependent formulation for $\sigma(M_r^h)$ brought the clustering of the mock closer to that of observations. 

$\sigma(M_r^h)$ ranging from $\sim$0.4 for the brightest galaxies and $\sim$1.2 for the faint end produced favourable clustering results in our analysis. The sigmoid shape of tanh(x) and the fact that tanh(x) is bound to ($-1,+1$) naturally brought us to the values $\alpha \sim 0.8$ \& $\beta \sim 0.4$.

When we first implemented the scatter using a standard, non-clipped, Gaussian, objects that started out with low luminosities overwhelmed the brightest population because of the comparatively large abundance of the low-luminosity objects.  This leads central galaxies to form a bimodal distribution at masses of friends-of-friends haloes with  $\log_{10}({M_{200, \textup{mean}}}/{h^{-1} M_\odot}) > 13$. This is a result of the fact that our method for adding scatter to \Mhr \ does not distinguish between satellite and central subhaloes. We tried a variety of modifications to our scatter method and found that clipping the Gaussian in our scatter at 2.5 $\sigma$ solved the problem of false central galaxy \Mhr \ bimodality.  We discuss this further in Section~\ref{sec:fate of centrals}.

Our definition of \zform \ as the subhalo attribute that connects a subhalo's colour to its history was inspired by \citet{Masaki2013a} and \citet{Yamamoto2015}; with a modification that our \zform \ is defined in terms of \vpeak , as opposed to \vmax .

To calculate our clustering results, we use the publicly available code corrfunc\footnote{\url{https://github.com/manodeep/Corrfunc}} \citep{Corrfunc1, Corrfunc2}.

\section{Properties of the Rosella catalogue}\label{sec:properties}

In this section, we examine the Rosella mock produced using the methodology introduced in Section~\ref{sec:methods}.  In Section~\ref{sec: LF}, we open with a discussion of the properties of the luminosity function of the galaxies in Rosella and discuss the brightness limits that it can potentially reach, followed by, in Section~\ref{sec:colour bimodality in Rosella}, the galaxy colour bimodality. Section~\ref{sec:fate of centrals} describes the impact that our model of luminosity scatter has on the distribution of central and satellite galaxies in the mock. Section~\ref{sec:clustering analysis} considers the clustering in our mock, with a comparison to previously published observational and simulated data.

\subsection{Rosella luminosity function and resolution}\label{sec: LF}

\begin{figure}
	\begin{center}
		\includegraphics[width=\columnwidth]{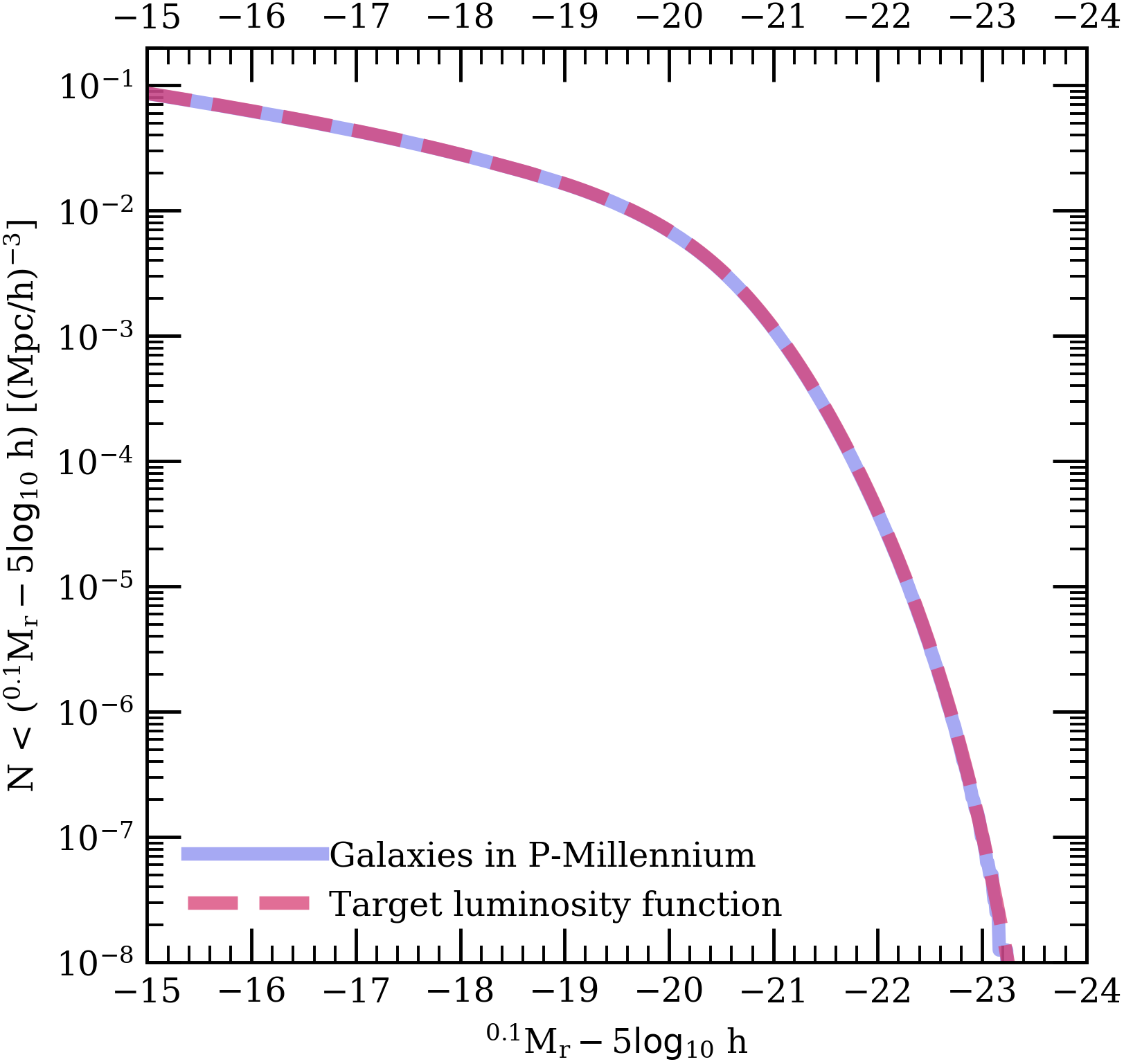} 
	\end{center}
\caption[Luminosity function of the mock catalogue.]{The \textit{r}-band cumulative luminosity function. The function for galaxies in the mock catalogue is plotted as a solid violet line. The pink dashed line is the target luminosity function based on SDSS and GAMA observations, taken from the fit provided in \citet{Smith2017}.}
	\label{fig:lum fn}
\end{figure}

\begin{figure}
\centering
\includegraphics[width=\columnwidth]{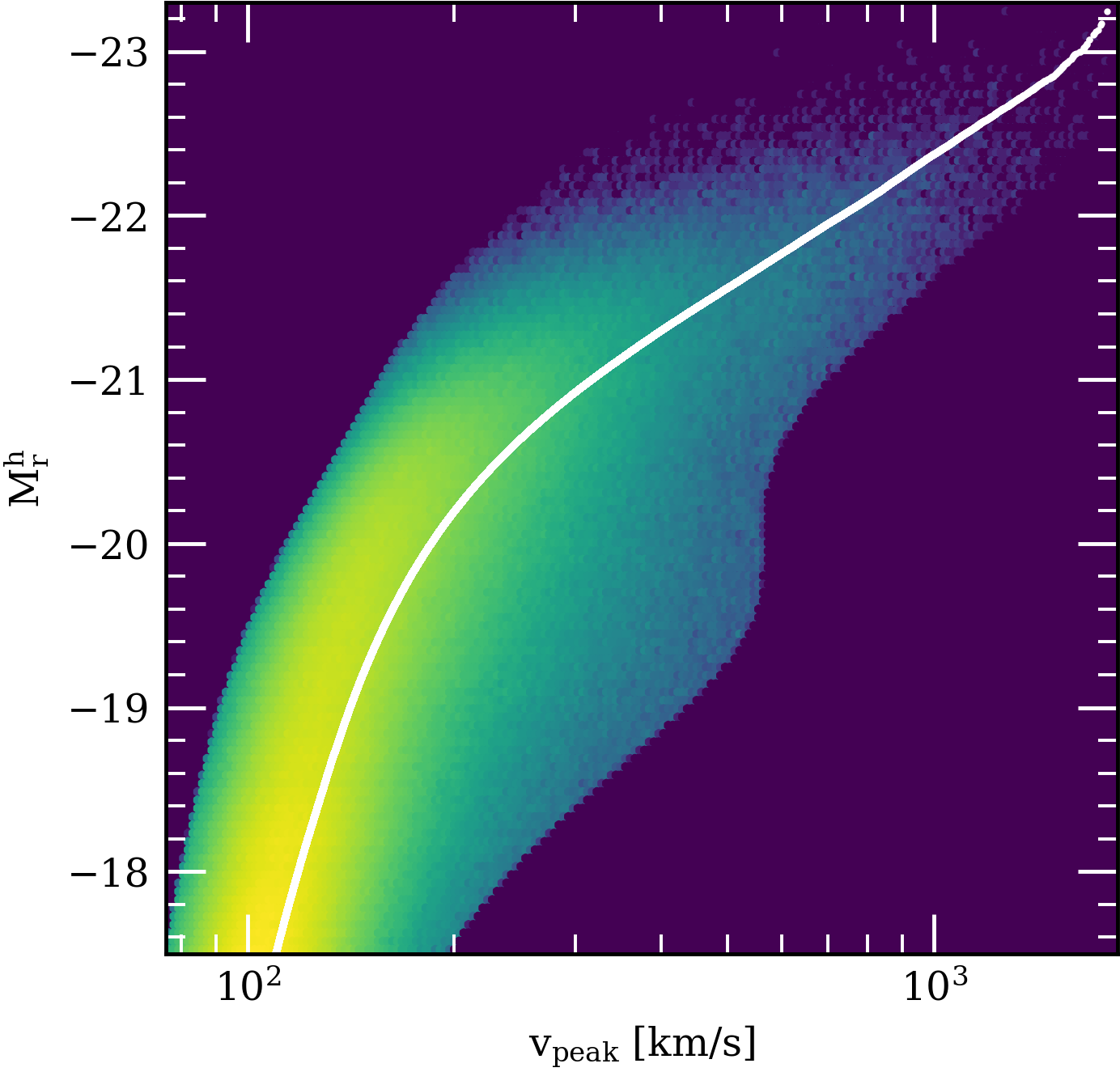}
\caption{Hexbin map of SHAM absolute magnitudes with scatter. The colour indicates the number of galaxies per hexagonal bin of given \Mhr \ and \vpeak \ values, plotted on a logarithmic scale, with purple indicating bins with zero galaxies and lime-green indicating bins with the most galaxies. The white line plotted on top of the hexbin map shows the \Mhr \ values assigned to P-Millennium subhaloes before the addition of scatter. The density of galaxies in the brightest yellow regions is about 5 orders of magnitude higher than the faintest nonzero blue regions.}
\label{fig:full Mhr vs vpeak}
\end{figure}

\begin{figure}
\centering
\includegraphics[width=\columnwidth]{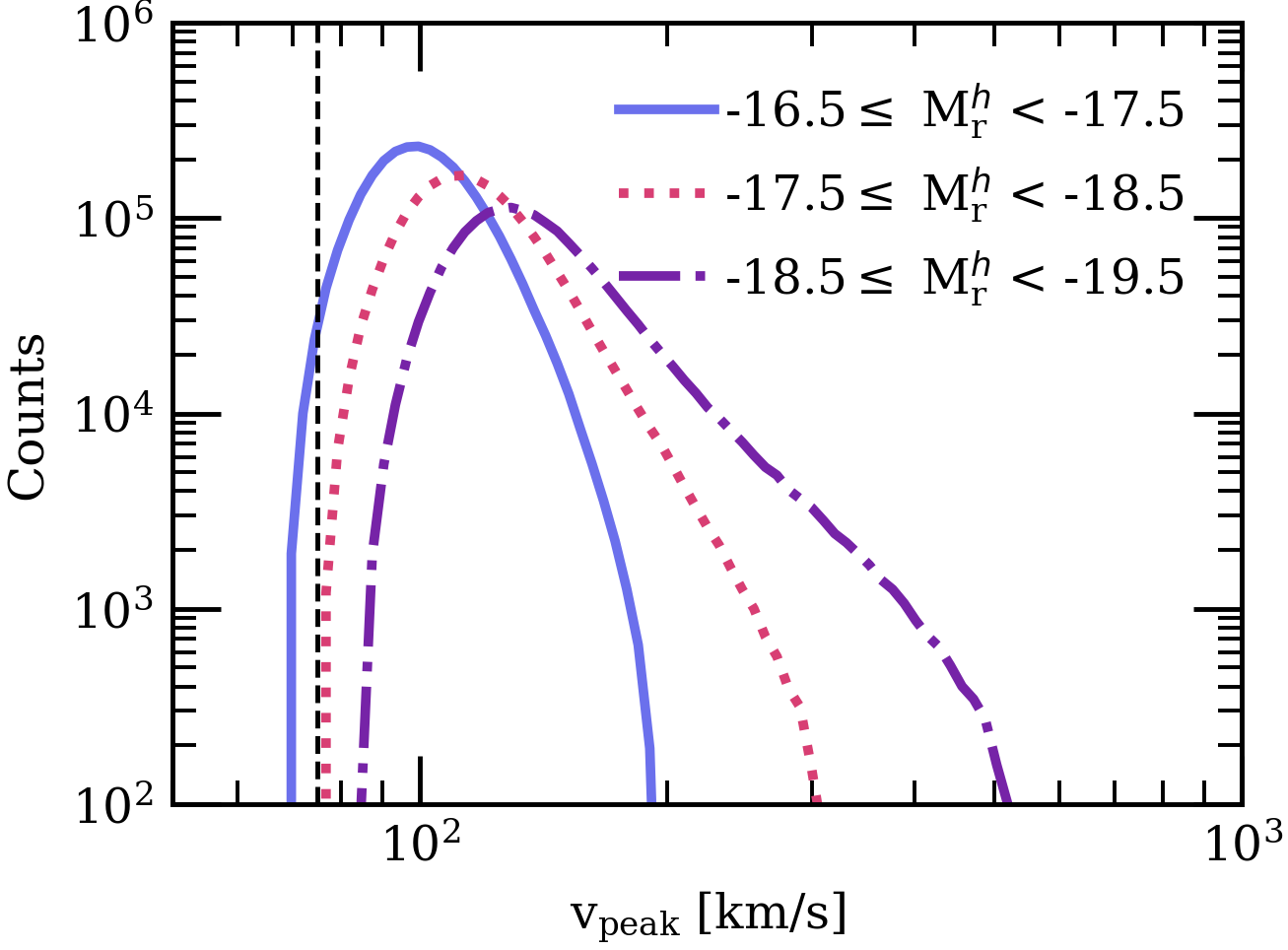}
\caption[Histograms of \vpeak \ in bins of \Mhr]{Histograms of \vpeak \ in bins of \Mhr, created by drawing \Mhr -limited samples, as indicated in the legend, from the full set of galaxies depicted in Fig.~\ref{fig:full Mhr vs vpeak}. The histograms cover the same set of 125 bins that cover the range $45$ km/s $<$ \vpeak \ $<2500$ km/s. The dashed black vertical line indicates the \vpeak $=75$ km/s boundary.}
\label{fig:vpeak histograms}
\end{figure}

By construction, the implementation of SHAM used here reproduces its target luminosity function. Fig.~\ref{fig:lum fn} demonstrates that the luminosity function produced in our mock exactly matches the cumulative galaxy luminosity function based on SDSS \citep{Blanton2003} and GAMA \citep{Loveday2012} data provided in \citet{Smith2017} to magnitudes at least as faint as \Mhr $=-15$. This, however, does not mean that the properties of the Rosella catalogue are converged at such faint magnitudes: these faint galaxies may reside in haloes of such low \vpeak\ as to be where the P-Millennium catalogue is incomplete. Moreover, to assign a colour to a Rosella galaxy, we need to have a reliable \zform \ for such haloes. Earlier in Fig.~\ref{fig:colour resolution}, we saw that we require \vpeak \ $> 75$~km/s for \zform \ to be well defined. Hence, to determine the magnitude limit down to which Rosella is complete and produces reliable colours, we need to identify the magnitude at which the mock galaxies reside only in haloes with \vpeak \ $> 75$~km/s. This is revealed in Figs.~\ref{fig:full Mhr vs vpeak} and~\ref{fig:vpeak histograms}.

Fig.~\ref{fig:full Mhr vs vpeak} shows the SHAM absolute magnitudes as a function of \vpeak\ before (white curve) and after (hexbin colour map) scatter has been added. Histograms through this distribution are shown for three magnitude bins in Fig.~\ref{fig:vpeak histograms}. From these, we see that the magnitude bin extending as faint as \Mhr$=-17.5$ tapers smoothly to zero above \vpeak $=  75$~km/s, indicating that our catalogue is complete to this magnitude limit.

The lower limit on the absolute magnitude that produces a complete sample of galaxies may vary if one were to add scatter that follows a functional form different from equation~\ref{eq:deltamr} or utilise a different set of parameters, or apply this method to a simulation other than P-Millennium.

\subsection{Galaxy colour bimodality}\label{sec:colour bimodality in Rosella}

\begin{figure}
\centering
\includegraphics[width=\columnwidth]{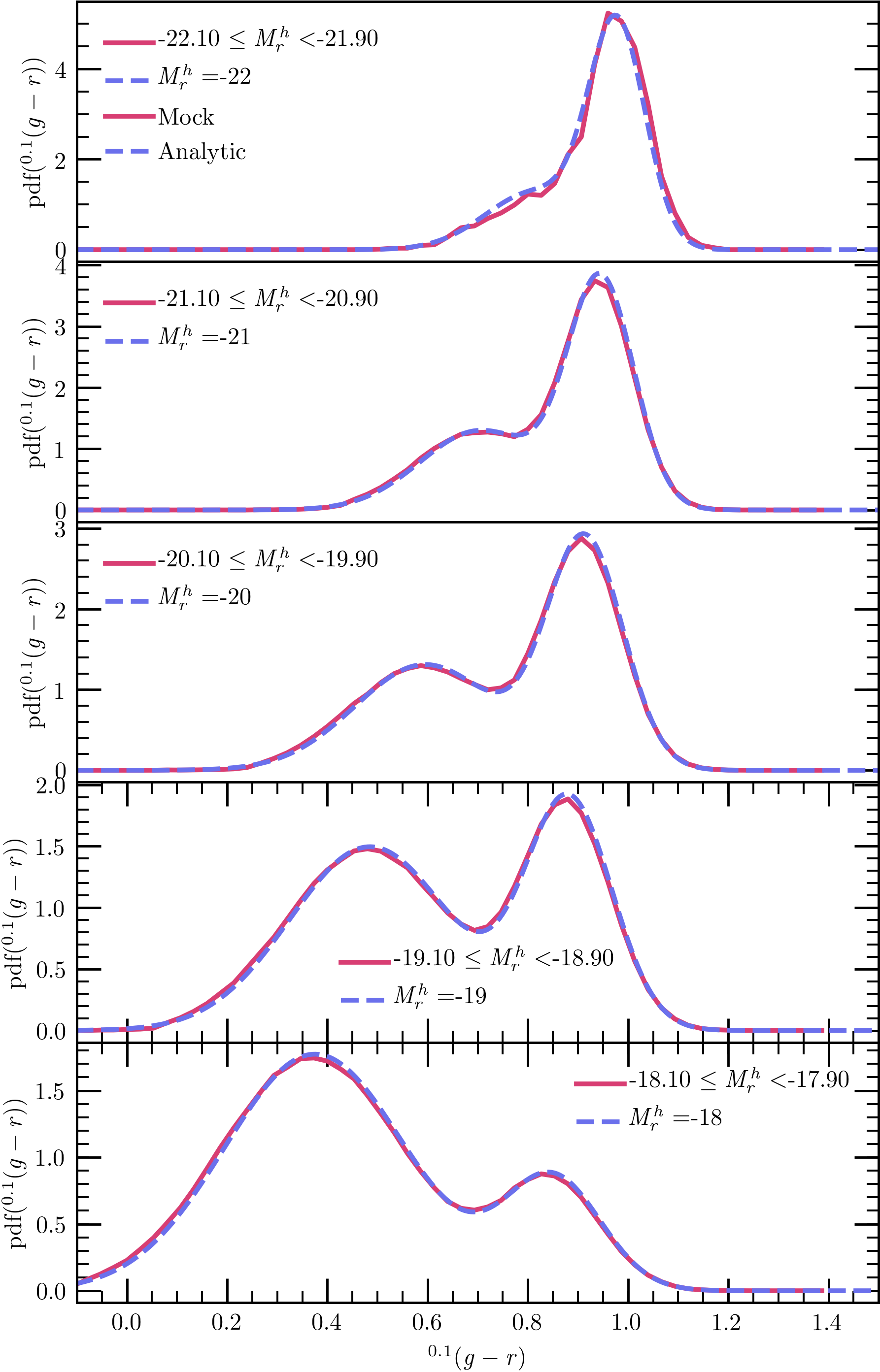}
\caption{Distribution of \gr \ values among Rosella galaxies. The red (solid) shows a normalised histogram of Rosella galaxies that fall in the range of \Mhr \ values indicated in the legend. The blue (dashed) line is the input function~\ref{eqn: colour cdf}, calculated at values of \Mhr \ indicated in the legend of each panel.}
\label{fig:analytic to data colour comparison}
\end{figure}

\begin{figure}
\centering
\includegraphics[width=\columnwidth]{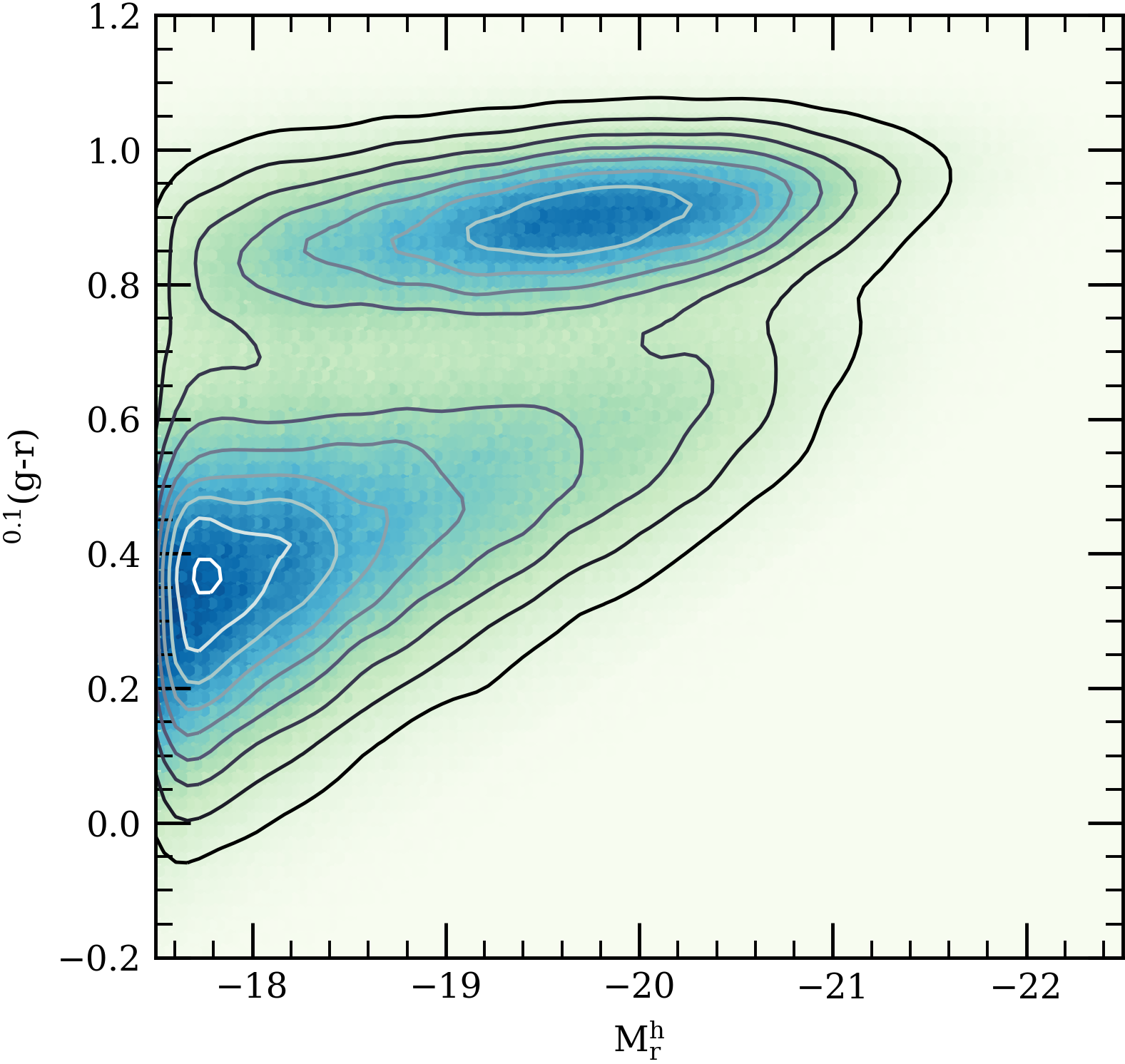}
\caption[Colour-magnitude diagram of Rosella galaxies]{Colour-magnitude diagram of Rosella galaxies as a hexbin map with contours. The  map shows the density of galaxies in hexagonal bins of \gr \ and \Mhr \ values. Fig.~\ref{fig:analytic to data colour comparison} shows slices through this distribution.}
\label{fig:colour-magnitude diagram}
\end{figure}

Fig.~\ref{fig:analytic to data colour comparison} shows histograms of \gr \ colour values in Rosella, along with curves produced by the analytic expressions generated at specific \Mhr \ values with equation~\ref{eqn: colour cdf}. By construction, we match the colour distributions in \citet{Smith2017}, which were designed to match SDSS and GAMA data. The resulting colour distributions are compared to observational GAMA data in figure~14 of \citet{Smith2017}. The histograms in Fig.~\ref{fig:analytic to data colour comparison} reveal a good match with the target colour distributions examined in figure~14 in \citet{Smith2017}, which, in turn, provide a good match to those of the SDSS and GAMA surveys \citep[e.g.][]{Baldry2004a}. Each histogram generally shows two major peaks with the blue being dominant for low luminosity and the red for high luminosity. The location of both peaks moves redward with increasing luminosity.  These distributions combine to produce the colour-magnitude diagram shown in Fig.~\ref{fig:colour-magnitude diagram}, whose morphology is akin to those shown in the literature \citep[e.g.][]{Baldry2004a}.

\subsection{Distribution of central and satellite galaxies}\label{sec:fate of centrals}

\begin{figure}
\centering
\includegraphics[width=\columnwidth]{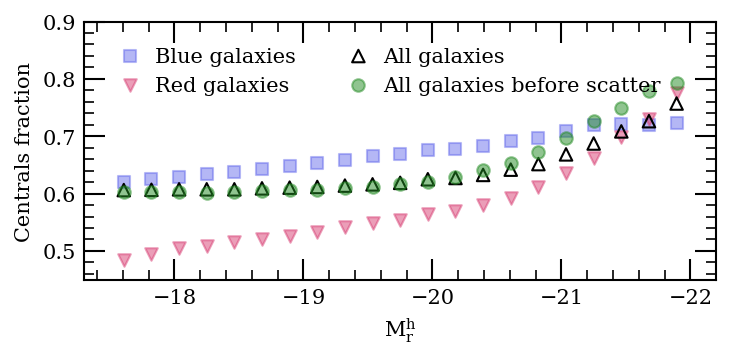}
\caption{Fraction of red (filled red triangles) or blue (blue squares) galaxies that are centrals given the galaxies' \Mhr; fraction of all galaxies that are central in Rosella with (empty triangles) and without (green circles) scatter as a function of the galaxies' \Mhr. Blue and red galaxy populations are defined in equation~\ref{eq: colour cut}.}
\label{fig: central fractions}
\end{figure}

The number of satellite galaxies as a fraction of the total galaxy population in Rosella varies with halo mass. In the top panel in Fig.~\ref{fig: central fractions}, one can see that galaxies that are assigned brighter absolute magnitudes with SHAM before scatter are preferentially central galaxies. In both cases of SHAM samples with and without scatter, the trend in the ratio of central galaxies to the total galaxy population tapers off to an almost constant rate of about 60\% between \Mhr \ $=-20$ and  \Mhr \ $=-17.5$. The scattered sample of SHAM, however, exhibits a lower fraction of satellites compared to the no-scatter sample at the bright end of the catalogue. This is the result of the scattering process moving the magnitudes of galaxies that start out in central subhaloes to satellite subhaloes.

Fig.~\ref{fig: central fractions} shows the fractions of blue and red galaxy populations that are central, given the galaxies' \Mhr. The nominal separation between ``red'' and ``blue'' galaxies is given by an expression introduced in \citet{Zehavi2005}:
\begin{equation}\label{eq: colour cut}
\grcut = 0.21 - 0.03 M_r^h.
\end{equation}
Galaxies whose \gr\  values are greater than this \gr\textsubscript{cut} are classified as ``red'', while the others are ``blue''.

 The trend in Fig.~\ref{fig: central fractions} demonstrates a steady increase in the fraction of central galaxies across the range of absolute magnitudes among Rosella galaxies in both red and blue galaxies. The blue population has a higher central galaxy fraction compared to the red population across all magnitudes, except for the brightest bins, with \Mhr $<\sim-21.5$.

\begin{figure}
\centering
\includegraphics[width=\columnwidth]{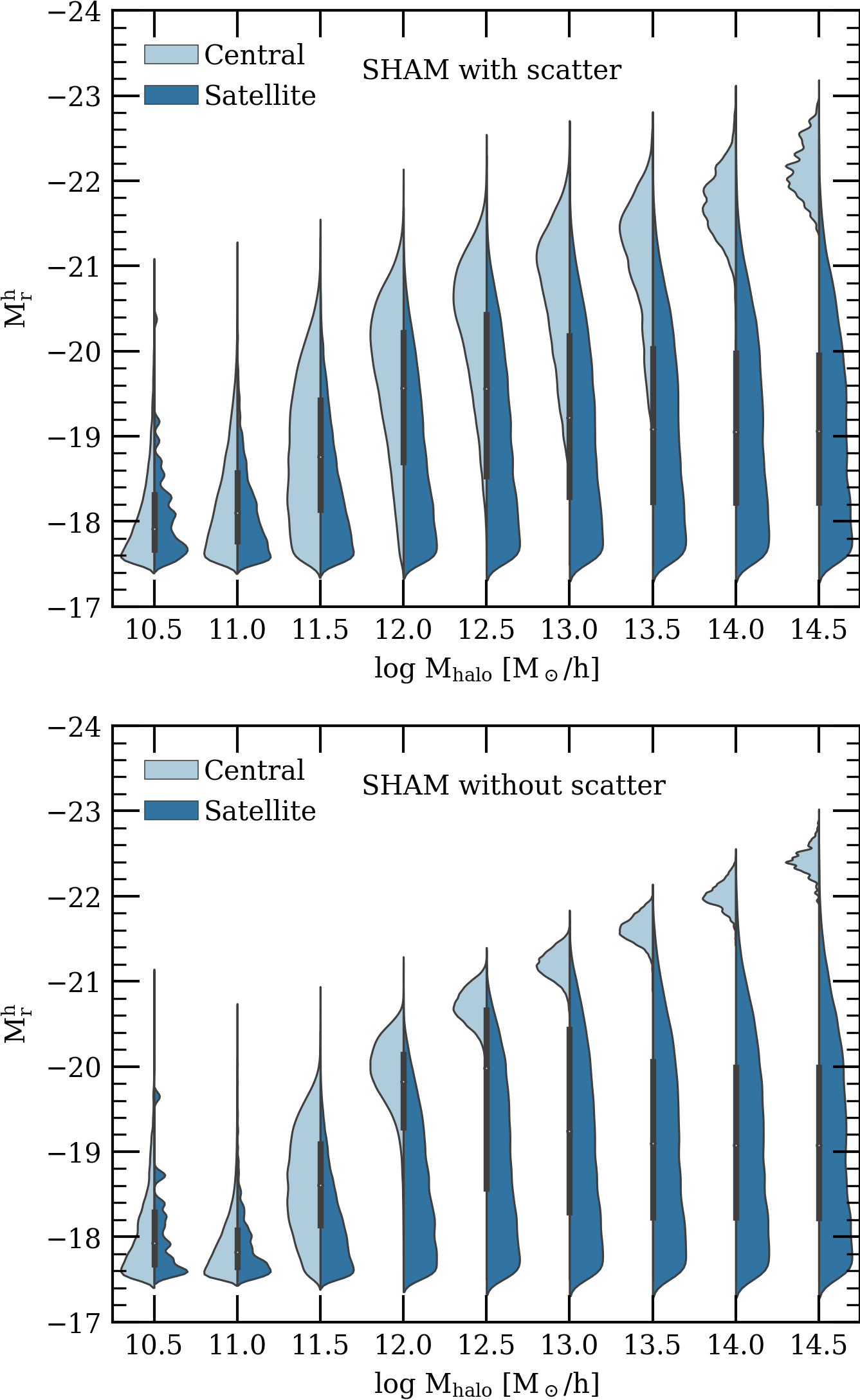}
\caption[Distribution of central and satellite galaxies in halo mass bins.]{Distribution of central and satellite galaxies in halo mass bins for a sample of Rosella galaxies with \Mhr $<-17.5$. The vertical axis shows \Mhr, and the horizontal axis represents the bins of host halo mass (M\textsubscript{200, mean}\footnotemark). The top panel shows the distributions of satellite (light blue) and central (dark blue) galaxies with respect to their \Mhr \ values in bins of halo mass in Rosella with scatter described in Section~\ref{sec: luminosity algorithm}. The bottom panel shows analogous distributions for a SHAM catalogue with no scatter. Kernel smoothing has been applied to these violin histograms, which creates the false illusion of data stretching to magnitudes fainter than \Mhr \ of $-17.5$. The plots are normalised in a way that lets all histograms have the same width to draw our attention to the distribution of galaxies along the \Mhr \ axis, and not to the relative sizes of these populations.}
\label{fig: central violins}
\end{figure}

Fig.~\ref{fig: central violins} shows the normalised distributions of central and satellite galaxies in bins of host halo mass\footnotetext{Host halo mass, M\textsubscript{200, mean}, is defined by the mass enclosed within a radius at which the mean interior matter overdensity is 200 times the mean density of the Universe.} of 0.5 dex width. We see that the no-scatter SHAM sample (bottom panel of Fig.~\ref{fig: central violins}) exhibits a clear and expected trend of the peak of the distribution of centrals in the catalogue moving to a brighter magnitude with increasing halo mass.

The top panel in Fig.~\ref{fig: central violins} shows that when we add scatter using the formulation in Section~\ref{sec: luminosity algorithm}, the distinct population of central galaxies is preserved. This result comes from trying different prescriptions for adding scatter to \Mhr , and was achieved when we combined the luminosity dependent scatter (equation 6) with a Gaussian distribution clipped at 2.5 $\sigma$.

\subsection{Real- and redshift-space clustering in Rosella}\label{sec:clustering analysis}

\begin{figure}
	\includegraphics[width=\columnwidth]{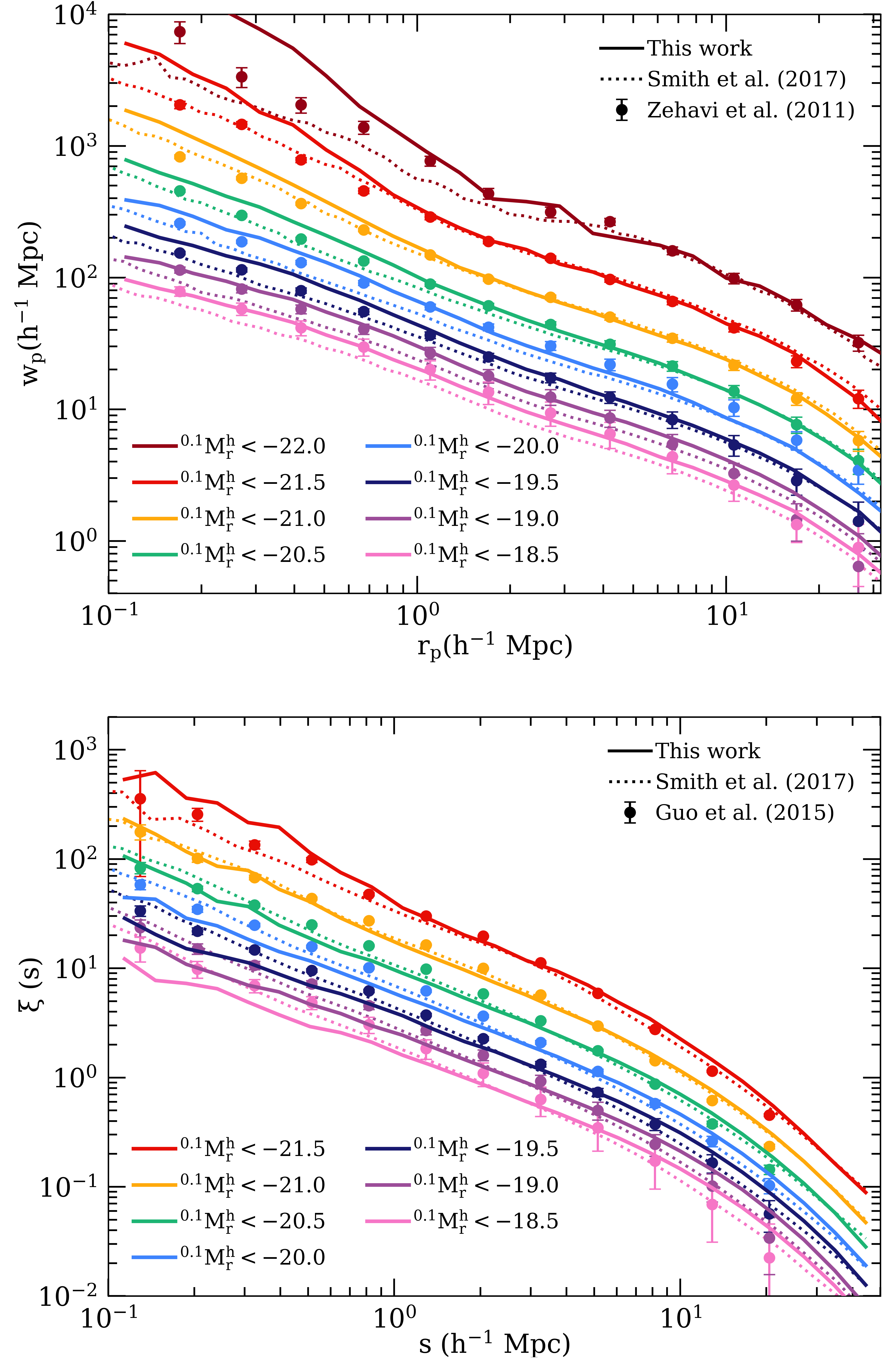}
    \caption{Projected correlation function (upper panel) and redshift-space correlation function (lower panel) for luminosity threshold galaxy samples. The solid lines show the clustering results of Rosella. The solid points with error bars represent clustering measurements using SDSS data from \citet{Zehavi2011} (upper panel) and \citet{Guo2015} (lower panel). The dotted lines show the corresponding correlation functions from the Millennium-XXL mock catalogue in \citet{Smith2017}. The results for each sample have been offset by successive intervals of 0.15 dex, starting at the \Mhr < $-20.5$ sample, with the faintest sample corresponding to the lowest curve in the graph, for clarity.}
    \label{fig:Lum-dep wp and xi}
\end{figure}

Studies of the clustering of early- and late-type galaxies, classified by spectral type, offer observational evidence of the dependence of the strength of galaxy clustering on morphology and luminosity. Observational evidence points to a trend in the spatial correlation function, where brighter galaxies are more clustered than their fainter counterparts \citep[e.g.][ and references therein]{Norberg2001, Zehavi2005}. Early studies of this phenomenon considered red and blue galaxies classified by spectral type, and observed that galaxies that belong to the ``early type'' population, which has been shown to be dominated by red and quenched galaxies, is more clustered than the ``late type'' population \citep[e.g.][ and references therein]{Norberg2001, Norberg2002, Zehavi2005}. The relatively high clustering of more luminous, redder galaxies, has led the luminous red galaxy (LRG) population to be a popular target sample for galaxy surveys that aim to study the large scale structure of the Universe \citep[e.g.][]{Eisenstein2005a, Eisenstein2005b}.

The luminosities and colours assigned to our high-fidelity mock offer a possibility of comparing the colour- and luminosity-dependent correlation functions to the trends observed in past surveys.
\subsubsection{Clustering as a function of luminosity}\label{sec:luminosity clustering}

Projected correlation functions of galaxies in Rosella are shown by the bold curves in the upper panel of Fig.~\ref{fig:Lum-dep wp and xi} for different luminosity threshold samples at \zapp 0.2. In the figure, we show the projected two point correlation functions (2PCF) calculated  using the publicly available code corrfunc \citep{Corrfunc1, Corrfunc2}. 

The samples presented in the upper panel of Fig.~\ref{fig:Lum-dep wp and xi} show the projected 2PCF in samples of galaxies with a faint limit on absolute magnitude (luminosity threshold). The sample cut-off limits have been chosen to make it possible to compare the clustering results of Rosella data to those of the HOD mock presented in \citet{Smith2017} and of the SDSS data presented in \citet{Zehavi2011}. It should be noted, however, that in addition to luminosity thresholds, the observed clustering of galaxies in \citet{Zehavi2011} and \citet{Smith2017} was measured for volume-limited samples. Each volume-limited sample covers a specific range of redshifts, and the range is wider for the bright samples. Rosella, on the other hand, is a single snapshot at \zapp 0.2, which may result in slight differences in the clustering of galaxies in Rosella and the SDSS data \citep{Zehavi2011} and MXXL mock \citep{Smith2017}. Considering that the SDSS and MXXL mock data do not cover the same redshift sample as Rosella, a more robust comparison of Rosella clustering to data would require detailed tuning of Rosella using data that is centered on \zapp 0.2, which will be available from DESI. 

While Rosella's projected 2PCF fits the SDSS data quite well on scales greater than $1\ h^{-1}$ Mpc, all but the two faintest samples exhibit clustering that appears to be slightly too high on small scales. We suspect that this might be a result of our SHAM model treating satellite and central galaxies in the same manner. Whether this feature of the model is compatible with quenched fraction estimates in \citet{Mandelbaum2016} is worth investigating in further work.

Additionally, the MXXL mock included galaxies assigned to haloes which were given random positions, corresponding to haloes below the mass resolution of the MXXL simulation. This random position assignment dilutes the clustering of MXXL galaxies slightly for faint galaxy samples, which explains why the clustering of the MXXL mock is low compared to Rosella for the \Mhr <$-18.5$ and \Mhr <$-19$ samples.

We have conducted the luminosity-dependent clustering analysis for a variety of models of scatter during the process of tuning our mock, presented in Section~\ref{sec:tuning}.

The lower panel in Fig.~\ref{fig:Lum-dep wp and xi} shows the redshift-space correlation function monopole for Rosella, compared to the redshift-space clustering of the mock presented in \citet{Smith2017}, as well as clustering of SDSS data from \citet{Guo2015}. We note a slight difference in shape and amplitude of the redshift-space 2PCF monopole between Rosella and SDSS. Addressing this difference would require further in depth analysis that we leave for a future work.

\subsubsection{Clustering as a function of colour}\label{sec: colour-dependent clustering}

\begin{figure}
\includegraphics[width=\columnwidth]{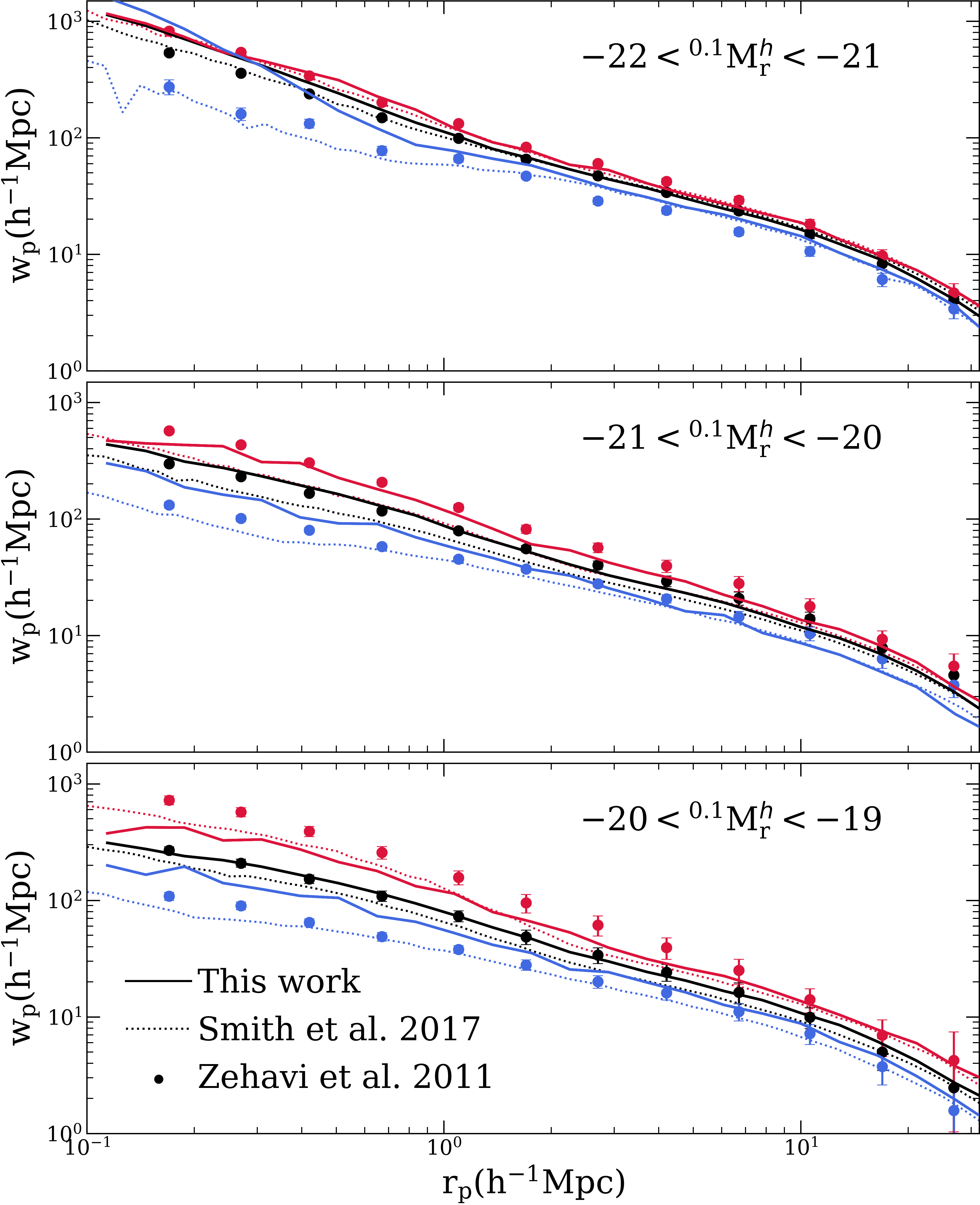}
\caption[Projected correlation function for red and blue galaxies.]{Projected correlation function for red and blue galaxies in bins defined by absolute magnitude. The clustering of Rosella galaxies is presented in bold lines. Clustering of low-redshift galaxies from the Millennium-XXL catalogue in \citet{Smith2017} is plotted in faint lines. The lines in each box correspond to bins defined by ranges of absolute magnitude, as indicated in the legend. Points with error bars correspond to the analysis of volume-limited samples of SDSS data in \citet{Zehavi2011}. The clustering of all galaxies in a sample is shown in black. Red and blue galaxy populations, defined by equation~\ref{eq: colour cut}, are plotted in red and blue colours, respectively.}
\label{fig:colour clustering}
\end{figure} 

Fig.~\ref{fig:colour clustering} shows the projected correlation function of Rosella galaxies separately for red and blue galaxy populations in bins of absolute magnitude. The same figure shows a comparison of our data to those presented in \citet{Smith2017} and \citet{Zehavi2011}, where red and blue samples are defined using the same colour cut as this work's, given by equation~\ref{eq: colour cut}.

For the purposes of analysis, the nominal separation between ``red'' and ``blue'' galaxies is given by equation~\ref{eq: colour cut}.  It should be noted that this expression, first introduced in \citet{Zehavi2005}, does not account for the fact that there is a continuum in galaxy colours, and instead serves as a tool for comparing colour-dependent clustering among different samples.

For SDSS, the clustering of the red galaxy population is stronger than that of the blue galaxy population. This effect is likely associated with the presence of red elliptical galaxies, which are more likely to reside in the more strongly biased  massive haloes \citep[e.g.][]{Eisenstein2005b}. As the samples get fainter, the strength of the colour dependence evidently increases for both the observational data and the galaxies presented in Rosella.

In summary, the comparison of Rosella galaxy clustering shows a favourable match to SDSS data, considering the differences that may arise from comparing Rosella's fixed-redshift sample to the volume-limited samples that cover ranges of redshifts in SDSS. This is, therefore, useful for developing analyses of DESI BGS.

\section{Conclusion}\label{sec:conclusion}
Modern galaxy surveys require realistic mock catalogues in order to test analysis tools, assess completeness, and determine error covariances in observed data. The mock catalogues can serve as the connector of quantities that the surveys observe, such as galaxy luminosities and positions, to quantities that are only available in simulations, including but not limited to host halo and subhalo masses, velocities, and halo assembly histories.

We have outlined a method for creating a mock galaxy catalogue that closely mimics data that will be observed in DESI's Bright Galaxy Survey  \citep{DESI2016, Ruiz2020}. The resulting mock, Rosella, provides the rest-frame \textit{r}-band absolute magnitudes, rest-frame \gr\  colours, 3D positions and velocities for galaxies inhabiting a volume of approximately (542 Mpc/h)\textsuperscript{3}, as well as the masses of their host haloes. 

The approach described here relies on SHAM with luminosity-dependent scatter to populate the P-Millennium N-body simulation with galaxies and assign them rest-frame absolute magnitudes in $r$-band, \Mhr. Due to our approach to adding scatter to the mock, Rosella preserves the target luminosity function by construction.
Our method also faithfully reproduces a specified redshift-dependent target distribution of \gr \ colours.  The colours it assigns are linked to the formation redshifts we determine by tracking the formation history of each individual subhalo. In correlating colour with formation time, we are following an approach similar to e.g. \citet{HearinWatson2013}.

As a reference mock, Rosella will be useful for fulfilling tasks that include analysing galaxy survey biases and calibrating approximate mocks that can scale up the galaxy population data in Rosella to meet volume and abundance requirements.

The mock presented here may be useful for low-redshift galaxy surveys that could benefit from a \zapp 0.2 reference mock. The method behind Rosella can further be used to generate galaxy catalogues at other redshifts. Should one need a lightcone catalogue with galaxies populated with the Rosella method, one could populate other snapshots in the P-Millennium simulation, and produce a lightcone from the resulting suite of reference mocks that correspond to fully populated boxes of the P-Millennium volume. The method used here can thus benefit any survey that probes volumes similar to those covered by P-Millennium.

Compared to the HOD-based mock presented in \citet{Smith2017}, the Rosella mock includes a greater degree of assembly bias by construction from the \vpeak -based SHAM method for luminosity assignment and a colour assignment method that relies on each galaxy's individual subhalo history. Rosella connects the simulation-only properties that are not directly observable, such as halo and subhalo mass, to directly observable quantities, \Mhr \ brightness and \gr \ colour. This opens the possibility of using Rosella and the method behind it to search for evidence of assembly bias in galaxy surveys that probe volumes similar to Rosella's.

We evaluate the closeness of the match between our mock and real data by comparing the luminosity- and colour-dependent clustering of our mock's galaxies against previously published clustering of similar galaxy populations in existing observational and mock data. Users of Rosella and its method may be interested in other summary statistics, e.g. redshift-space distortions. 

The tuning of the mock for specific scientific goals may adjust the choice of free parameters in the creation of our data, such as the functional form and parameters in the luminosity-dependent scatter added to the \Mhr \ data, as well as the definition of \zform. While we have considered two values of $f$ in relating subhalo \vmax \ histories to \zform \ and found that the two options did not have a significant effect on colour-dependent clustering, other formulations of \zform \ might be possible and may suit specific scentific goals.

To put constraints on cosmological parameters using Rosella's linking of observable and simulation-based galaxy qualities, such as luminosity and halo mass, error covariances need to be determined. This requires the use of many mock catalogues, of the order of up to $10^4$ and greater \citep[e.g.][]{WhiteTinker2014, Kitaura2016, Quijote2019}. This can be achieved by calibrating fast mock generation methods using the reference mock presented here and, potentially, doing so at a variety of redshifts by applying the Rosella method to a variety of P-Millennium snapshots.

\subsection*{Acknowledgements}

SS has received funding from the U.S.-U.K. Fulbright Commission and the Gruber Science Fellowship. PN \&  SMC acknowledge the support of the Science and Technology Facilities Council (ST/L00075X/1 and ST/P000541/1).

The authors acknowledge Alex Smith, Jeremy Tinker and Eduardo Rozo for insightful comments.  The authors thank the groups behind \citet{Guo2015}, \citet{Smith2017}, and \citet{Zehavi2011} for providing data points from their publications.

This research is supported by the Director, Office of Science, Office of High Energy Physics of the U.S. Department of Energy under Contract No. DE--AC02--05CH1123, and by the National Energy Research Scientific Computing Center, a DOE Office of Science User Facility under the same contract; additional support for DESI is provided by the U.S. National Science Foundation, Division of Astronomical Sciences under Contract No. AST-0950945 to the NSF's National Optical-Infrared Astronomy Research Laboratory; the Science and Technologies Facilities Council of the United Kingdom; the Gordon and Betty Moore Foundation; the Heising-Simons Foundation; the French Alternative Energies and Atomic Energy Commission (CEA); the National Council of Science and Technology of Mexico; the Ministry of Economy of Spain, and by the DESI Member Institutions.  The authors are honored to be permitted to conduct astronomical research on Iolkam Du'ag (Kitt Peak), a mountain with particular significance to the Tohono O'odham Nation.

This work used the COSMA Data Centric system at Durham University, operated by the Institute for Computational Cosmology on behalf of the STFC DiRAC HPC Facility (\url{www.dirac.ac.uk}).

\subsection*{Data Availability}
The data underlying this article were accessed from the COSMA Data Centric system at Durham University (\url{www.dirac.ac.uk}). The derived data generated in this research will be shared on reasonable request to the corresponding author.


\bibliographystyle{mnras}
\bibliography{rosella_bib}




\bsp	
\label{lastpage}
\end{document}